\documentclass[12pt,preprint]{aastex}
\newcommand{\mic}{\hbox{$\mu$m}}      	       % \mic 
\newcommand{\ergsec}{\mbox{erg s$^{-1}$}}      	 % \erg/s 
\newcommand{\kms}{\mbox{km s$^{-1}$}}      	 % \km/s
\newcommand{\cms}{\mbox{cm s$^{-1}$}}      	 % \km/s
\newcommand{\thC}{\mbox{$\theta^1$Ori~C}}  	% \theta 1 Ori C
\newcommand{\Lya}{\mbox{Ly-$\alpha$}}
\received{30 November 2001}
\accepted{1 July 2002}
\slugcomment{%DRAFT: DO NOT PHOTOCOPY
To appear on Ap.J, v578 n2 October 20, 2002.}

\lefthead{Robberto et al.}
\righthead{SED of YSOs in HII region}

\begin{document}

\title{The Infrared Emission of Circumstellar Envelopes, Dark Silhouettes and Photoionized Disks in HII Regions}

\author{M. Robberto\altaffilmark{1}, S. V. W. Beckwith, N. Panagia\altaffilmark{1}}  
\affil{Space Telescope Science Institute, 3700 San Martin Drive, Baltimore, MD 21218, USA}
\email{robberto@stsci.edu, svwb@stsci.edu, panagia@stsci.edu}
\altaffiltext{1}{On assignment from Research and Science Department of ESA}

% The abstract environment prints out the receipt and acceptance dates
% if they are relevant for the journal style.  For the aasms style, they
% will print out as horizontal rules for the editorial staff to type
% on, so long as the author does not include \received and \accepted
% commands.  This should not be done, since \received and \accepted dates
% are not known to the author.

\begin{abstract}
We have modeled the infrared (IR) Spectral Energy Distribution (SED) 
of circumstellar disks embedded in a HII region and photoevaporated 
by the external ultraviolet radiation. The model applies to the 
photoevaporated disks (proplyds) in the Orion Nebula, most of them 
illuminated by the O6.5 star \thC.  First we calculate the IR emission
of a Pre-Main-Sequence star surrounded by a dusty globule that is immersed 
within an HII region.
The globule is assumed to be spherical, homogeneous, optically thin at IR wavelengths 
and photoevaporated according to the Dyson (1968) model.
Second, we consider the IR emission of a disk directly exposed to the nebular environment. 
The reprocessing disk is passive and treated according to the Chiang and Goldreich 
(1997, CG97) model. We improve over the CG97 treatment by tracing 
the propagation of the various radiative fluxes (from the star exciting the 
HII region, nebular, and grazing from the disk central star) through 
the disk superheated atmosphere. Since the opposite disk sides receive different
amounts of radiation, the flaring angle and the surface temperature distributions
are different, resulting in well distinguished SEDs for the two disk faces.
% and possibly disk warping. 
Finally, we combine the globule and disk models to estimate
the IR emission of proplyds. The energy input from the central star and the
nebular environment increase the disk flaring angle, and therefore also the amount of 
stellar radiation intercepted by the disk. 
%due to the dust in the evaporated envelope and in the disk's interior.
The relative intensity of the disk vs. envelope emission
varies with the tilt angle relative to the directions of \thC\, and 
of the Earth. We explore the dependence of the SEDs upon the tilt angle
with respect to the Earth, the distance from \thC, the size on the envelope, the inner
disk radius and the temperature of the central star.
The resulting SEDs
are characterized by a broad peak of emission at 30-60\,\micron\, and are
in general significantly different from those of isolated disks 
in low-mass star forming regions like Taurus-Auriga.  
Our model indicates that in the presence of 
an external radiation field, relatively evolved "Class 2" objects 
may display a SED peaking at mid-IR and far-IR wavelengths. Also,
the model can account for  the strong mid-IR excess we have recently detected 
at 10\,\micron\, from embedded disks in the Orion Nebula. 
\end{abstract}

% The different journals have different requirements for keywords.  The
% keywords.apj file, found on aas.org in the pubs/aastex-misc directory, 
% contains a list of keywords used with the ApJ and Letters.  These are 
% usually assigned by the editor, but authors may include them in their 
% manuscripts if they wish. 


% That's it for the front matter.  On to the main  body of the paper.
% We'll only put in tutorial remarks at the beginning of each section
% so you can see entire sections together.

% In the first two sections, you should notice the use of the LaTeX \cite
% command to identify citations.  The citations are tied to the
% reference list via symbolic KEYs.  We have chosen the first three
% characters of the first author's name plus the last two numeral of the
% year of publication.  The corresponding reference has a \bibitem
% command in the reference list below.
%
% Please see the AASTeX manual for a more complete discussion on how to make
% \cite-\bibitem work for you.   

\section{Introduction}

Circumstellar disks are thought to be the birthsites of planetary systems, and
it is primarily for this reason that their properties have been studied extensively 
since the
early 1980's (Shu, Adams, and Lizano 1987; Beckwith \& Sargent 1996; Hartmann
1998).  The discovery of several
tens of extra-solar planetary systems (e.g. Marcy, Cochran \& Mayor 2000) added
weight to the argument that disks create planets, although there is some debate
about how often planetary systems really occur around stars.  Nevertheless, an
understanding of the evolutionary properties of disks around young stars is
thought to be an essential part of the understanding of our own origins.

Disks can also affect the early evolution of stars themselves.  Several authors
have suggested that stellar rotation is locked to the orbital rates of the inner
disks at early times allowing the disk to regulate the stellar angular momentum
(Edwards et al. 1993, Choi \& Herbst 1996), although this idea is still
controversial (Stassun et al. 1999, Rebull 2001).  
Typical disk masses and accretion rates appear to be too small
to greatly change the mass of the stars through accretion (Palla \& Stahler, 1999).  
A number of observable characteristics of young star/disk systems
play an important role in determining the stellar properties
%rotation, for example, so 
and therefore 
understanding even subtle changes is important to ensure
that the stellar properties are correctly interpreted.

Most of the stars in the Galaxy are thought to be born in dense OB clusters such
as the Orion Nebula (Bally et al. 1998a; McCaughrean \& Stauffer 1994).  However,
our understanding of circumstellar disks comes mainly from studies of nearby dark
clouds such as the Taurus/Auriga complex, and it is not clear that the extant
knowledge of circumstellar disks applies to the majority of young star/disk
systems in the Galaxy.  The study of the disks in Orion presents several problems
that has slowed progress: they are seen against the Orion HII region, a bright,
highly nonuniform source of radiation at wavelengths from the ultraviolet through
millimeter; they appear to be fainter on average at millimeter wavelengths
(Bally et al. 1998b); and the Orion nebula is three times
farther from Earth than the well-studied dark clouds.
Many of the disks are
embedded within the HII region, meaning that they are surrounded by ionized halos of
gas and dust that complicate the observation of disk properties alone.

To address these difficulties, we present in this paper a series of models of the
emission from disks embedded within the Orion HII region, under the influence of other radiation sources in addition to the stars at their center. 
Section 2 is a discussion of various sources of 
radiation in the HII region to prepare for
the calculations of emission from discrete sources.  In the following sections,
we show how the different radiation affects the IR emission arising from
three type of sources:
a spherically symmetric, dusty globule surrounding a star (section 3),
a star/disk system directly exposed to the ionized environment (section 4), and finally a
star/disk system surrounded by a dusty globule (section 5).  Section 6 contains a discussion of our findings with an exploration of the critical parameters, comparison with observations, and remarks on the limit of our treatment.

\section{General parameters of the HII region}

We assume the HII region is spherical, isothermal, and has uniform density.
For simplicity, we consider the brightest star at its center
as the only source of  ionizing photons, neglecting the radiation emitted by other stars within 
or around the nebula. This approximation is appropriate
for the Trapezium cluster, where \thC\, supplies at least 80\% of the Lyman continuum 
luminosity: $L_{Lyc}\approx 8\times 10^4\,L_{\sun}$
versus a total luminosity of $\sim 10^5\,L_{\sun}$ for the entire cluster (Bally et al. 1998a). 
We assume the HII region is radiation bounded with Str\"omgren radius, 
$R_{HII}$, i.e. all the Lyman-continuum radiation is absorbed within the nebula. 
We use for the HII region the typical parameters of the Orion Nebula,
with a Str\"omgren radius $R_{HII} = 1 \times 10^{18}$\,cm and a distance $D = 450$\,pc.
%For brievity, we will refer to the ionizing central star as to \thC.

To estimate the relative importance of the various dust heating mechanisms,
it is appropriate to distinguish between the stellar, $F_s$, and the nebular, $F_n$, radiative fluxes (Panagia 1974, Natta \& Panagia 1976). 
The stellar flux is comprised of ionizing, $F_s^{EUV}$, and nonionizing, $F_s^{FUV}$, radiation. The former maintains the ionization balance and is mostly absorbed in the vicinity of the Str\"omgren boundary. Once absorbed, it is transformed into nebular radiation, 
$F_n$, that fills the HII region uniformly. $F_n$ can be divided into the resonantly scattered \Lya\,  radiation, $F_n^{Ly-\alpha}$, and the rest, e.g. Balmer continuum and lines, forbidden lines from various
atomic species, continuum emission, etc., in short all referred to hereafter as nebular radiation, and indicated by $F_n^{other}$. 

If $\rho$ denotes the fraction of total luminosity, $L_s$, emitted by the central star 
in the Lyman continuum (Extreme Ultraviolet -- EUV; $h\nu >h\nu_{LyC}=13.6$\,eV), the ionizing flux at a distance, $d$, is:
\begin{equation}
F_s ^{EUV} (d) = \rho \frac{{L_s }}
{{4\pi d^2 }}e^{ - \tau_{EUV}(d) }, 
\end{equation}
\noindent
where $\tau_{EUV}(d)$ is the radial optical depth at the effective wavelength of the ionizing radiation,
$\lambda_{EUV}$ ($\sim 570$\AA\, for \thC). 
The non-ionizing stellar radiation (Far Ultraviolet -- FUV; 
6\,eV$<h\nu <13.6$\,eV) is
\begin{equation}
F_s ^{FUV} (d) = (1 - \rho )\frac{{L_s }}
{{4\pi d^2 }}e^{ - \tau_{FUV}(d) } 
%\eqno{(2)}
\end{equation}
\noindent
with an effective wavelength, $\lambda_{FUV} \simeq 1200$\AA\, (Panagia 1974). 

With the exception of the dusty globules and disks, discussed later in this paper, 
we assume that the amount of dust mixed with the ionized gas is negligible, so that both the EUV
and FUV fluxes cross the HII region almost unaffected by dust. 

Recombination theory predicts that as long as the electron density is $<< 10^4$\,cm$^{-3}$ a fraction $\simeq 2/3$ of the Ly-continuum photons are eventually transformed into \Lya\, photons at $\lambda_{Ly-\alpha}=1216$\,\AA\, through the 2p-1s transition, the remaining decaying to 1s via two-photon emission from the 2s level (e.g. Osterbrock 1989). The generation rate of Ly-$\alpha$ photons is therefore
\begin{equation}
N^{Ly-\alpha}  \cong \frac{2}{3} \frac{\rho L_s}{h\nu_{LyC}} ,
\end{equation}
and the corresponding luminosity is
\begin{equation}
L_n ^{Ly-\alpha }  \cong \frac{2}
{3}\rho L_s \left( {\frac{{\lambda _{LyC} }}
{{\lambda _{Ly - \alpha } }}} \right) \simeq \frac{1} 
{3}\rho L_s ,  
\end{equation}
where $\lambda_{LyC}=c/\nu_{LyC}=912$\,\AA.

For the ionizing star we 
adopt the parameters of an O6.5V star: $L_s  = 6 \times 10^{38} \ergsec$  and $\rho=0.4$ (Panagia 1973).
%We also assume $\alpha^\prime=3/2$.

For the remaining nebular radiation, 
\begin{equation}
L_n^{other}  = \rho L_s  - L_n ^{Ly - \alpha }  \simeq \frac{2}
{3}\rho L_s . 
\end{equation}

The energy density of the nebular radiation is obtained by multiplying the luminosity by the crossing time, 
$R_{HII}/c$, and dividing by the nebular volume, $V_{neb}=(4/3)\pi R_{HII}^3$. 
In the case of the \Lya\, radiation, the photons are resonantly scattered many times 
within the nebula before being absorbed. 
Numerical models (Panagia 1978, Hummer \& Kunasz 1980) indicate that whereas \Lya\, photons undergo approximately $10^5$ scatterings before reaching the boundary of an HII region, the space crossed between most of the scatterings is very short, so that a \Lya\, photon
reaches the boundary after having traveled on average a path length about $q \sim 10 - 20$ times the radius of the HII region. Therefore,
the absorption of \Lya\, photons by dust is about $q$  times more efficient 
than it is for non-resonant radiation with similar frequency. 
From the point of view of the dust heating by \Lya\, radiation, the factor,
$q$, can be treated either as an apparent increase of the \Lya\, energy density seen by the dust,
\begin{equation}
U_n ^{Ly - \alpha }  = \frac{{q\rho L_s }}{{4\pi c R_{HII} ^2 }},
%\eqno{(3)}
\end{equation}
or as an increase of the dust absorption efficiency $Q_{abs}$ at the $\Lya$ wavelength,
\begin{equation}
Q_{abs}(\Lya) = q Q_{abs}(UV).
\end{equation}
This second approach is more correct, since the apparent increase of the energy density 
holds only as long as \Lya\, photons 
are present, i.e. inside a dust-depleted HII region and at the outer edges 
of any embedded globule, disk, etc. In these objects, \Lya\, 
photons will cause an increase of the surface temperature, but at an optical depth to UV radiation 
$\tau\simeq 1$ they will be entirely absorbed.
Equation~(7) 
%provides the most convenient approach to treat absorption of resonantly trapped \Lya\, photons, which we have 
has been adopted in the rest of this paper.

Concerning the nebular radiation, one has
\begin{equation}
U_n^{other}  = \frac{{2\rho L_s }} {{4\pi c R_{HII} ^2 }}.
%\eqno{(4)}
\end{equation}

\section{Circumstellar globules}
\subsection{Neutral flow}
The ionized globules in the central part of the Orion nebula are typically shaped like 
tear-drops with the tail pointing away from \thC. For simplicity, 
we treat them as spheres of neutral gas surrounded by 
a shell that is photoionized on the hemisphere facing the ionizing star.
Johnstone, Hollenbach \& Bally~(1998) and St\"orzer \& Hollenbach~(1999) have modeled
the structure of the globules in terms of photoevaporated winds. They showed that
two different outflow regimes may develop: EUV-dominated 
and FUV-dominated flows. The former are produced when 
the ionizing radiation reaches the disk surface, typically 
in the vicinity ($d\lesssim10^{17}$\,cm) of the ionizing star, and/or in the very early phases of disk exposure
to the EUV radiation.
FUV-dominated flows are more common and occur when the EUV flux is absorbed at large distances 
from the disk surface. In this case, the disk photoevaporation is driven by the FUV radiation.
The photodissociation region (PDR) that develops  between the disk 
surface and the ionization front may extend
hundreds of astronomical units, or a few disk radii ($\sim 100$\,AU) from the disk.
Typical solutions of the flow equations indicate that 
in EUV-dominated flows the density remains approximately constant, whereas
in FUV-dominated flows the expanding neutral flow has nearly constant 
velocity until it reaches an isothermal shock, which separates the inner region 
with density $\propto r^{-2}$ 
from the isobaric upstream region where the density is approximately constant. 

St\"orzer \& Hollenbach (1999) have 
shown that in most FUV-dominated cases the shock front occurs much 
closer to the central star than to the ionization front. The region 
with constant density therefore fills most of the globule's volume. 
The dense inner flow may still dominate the radial column density, but
it is generated in the outer disk regions, where the thermal 
supersonic neutral wind at 2-6\,\kms\, exceeds the escape velocity 
from the rotating disk. The neutral outflow will be initially cylindrical by symmetry, while
a spherical flow
develops only at distances of the order of the disk size. The star should 
therefore remain in a cavity relatively free of dust, possibly dominated by the presence 
of collimated jets (Bally, O'Dell, \& McCaughrean, 2000). 
In conclusion, both EUV- and FUV- dominated flow models predict the 
presence of broad regions around the central star with uniform density. We shall 
therefore consider only FUV-dominated flows with constant density. Our neutral region
spans from the ionization radius $R_i$ down the an inner radius $R_0$, encircling
a dust-free cavity set by the dust 
evaporation temperature, $T_e=1,500$\,K. We shall refer to this structure 
as a ``neutral globule''. 

The density within the proplyds is regulated by the dust opacity to the FUV flux 
(Bally et al. 1998a).
%, because
If the density - and therefore, opacity - increases, the number of 
FUV photons reaching the disk diminishes. The smaller number of FUV photons
causes the evaporation rate to drop
with a consequent decrease of the density (opacity) in the globule. More FUV photons will 
then be able to reach the disk, keeping the system in equilibrium through negative feedback.
Both Johnstone et al. 
(1998) and St\"orzer \& Hollenbach (1999) estimated the column 
density of the neutral medium, $N(H)\simeq 10^{21}$\,cm$^{-2}$, corresponding to 
$A_V\simeq 0.5$ assuming average ISM opacity. This column density 
translates into a radial optical depth $\tau_{R_i} \simeq 1-2$ in the UV (both EUV and FUV), the exact value depending 
on the dust properties.  In the next section, we shall see how $\tau_{R_i}$ is directly related to the ionizing flux and
to the radius of the globule.

\subsection{Photo-ionized wind}
Behind the ionization front,
the gas flow is approximately spherically symmetric. 
At large distances from the globule, 
the interaction with the wind from the ionizing star will
create the characteristic stationary arcs of [OIII]+H$\alpha$ emission 
seen around several sources in the inner core of the Orion nebula (Bally et al. 1998a, Bally et al. 2000). 
Dust is trapped within the photoevaporating flow, as demonstrated by the arcs closer to \thC, which are prominent at 10\mic. In fact, the Ney-Allen nebula is mostly due to
the optically thin emission of dust in these arcs (Hayward, Houck, \& Miles 1994). 

Dyson (1968) first studied the evaporation of dense neutral globules in HII regions 
as a mechanism to provide high turbulence to the ionized medium. In Dyson's model, 
the globules are self gravitating spherical condensations of neutral hydrogen. Under the influence of the ionizing radiation, the ionized material at the globule surface 
streams out at supersonic velocities, injecting 
momentum and energy into the HII region. 
In this model, the spherically symmetric, steady flow has a solution given by
\begin{equation}
\frac{r}{R_i}=\sqrt{\frac {c_{II}} {v(r)} } 
                 exp{\left[ \frac  {v(r)^2-c_{II}^2}  {4c_{II}^2} \right] }
\end{equation}
and
\begin{equation}
\frac{n(r)}{n_i}=exp{\left[ \frac {c_{II}^2-v(r)^2} {2c_{II}^2} \right] },
\end{equation}
where $r$ is the distance from the center of the globule, $n(r)$ and $v(r)$ are the 
flow density and velocity, and $n_i$ is the atomic
density at $R_i$. 

We assume $c_{II}=10^6\,\cms$ for the sound speed in the ionized gas.
The Dyson solution neglects the role of gravity. In our case, gravity is
negligible at distances from a central star $r \ga 100$\,AU, since the escape velocity, $v_e=\sqrt{GM_\star/r}\simeq 1\,\kms$, for a central star of mass $M_\star = 1\,M_\odot$ is significantly lower than $c_{II}$ (see also Henney et al. 1996, Johnstone et al. 1998). In any case, the emitted IR spectrum is almost insensitive to the details of the density and velocity profiles.
  
In the expanding ionized flow, the neutral hydrogen and the grains compete for the 
absorption of the ionizing photons. The effects of dust in the 
photoevaporation flow has been studied by Pastor, Cant\'o, \& Rodriguez (1991),
and we refer the reader to their paper for the details. The optical depth within the globule is given by 
%Of particular interest here is 
their Equation~(12):
\begin{equation}
F=n_i c_{II} e^{\tau_d^\infty}+\alpha_B n_i^2 R_i b(\tau_d^0),
\end{equation}
valid in the plane-parallel approximation,
relating the flux of ultraviolet photons entering the globule, $F$, to the opacity from the ionization front to infinity, $\tau_d^\infty,$ to
%the radial opacity of a homogeneous globule with density $n_i$, 
$\tau_d^0 = \sigma_d n_i R_i$. 
Here, $\alpha_B$ is the recombination coefficient to the excited levels of the hydrogen,
and $\sigma_d=2\times10^{-21}$\,cm$^{-2}$ is the grain cross-section per hydrogen nucleus to the 
EUV radiation. The function, $b(\tau_d^0)$, has been tabulated by Pastor et al. (1991) for the Dyson's density profile. 
When $\tau_d^0\le 10$, one can approximate
\begin{equation}
b(\tau_d^0)=0.12e^{0.37\tau_d^0}.
\end{equation}
In this case (Pastor et al 1991):
\begin{equation}
\tau_d^\infty=0.45\tau_d^0.
\end{equation}

Figure~\ref{Pastor91EQ13_v3} shows how the optical depth, $\tau_d^0$, depends on the 
ionizing flux (i.e. distance from the ionizing star) and on the radius of the ionization front (i.e. 
globule's size) through Equations (11)-(13). 
In general, $\tau_d^0$ increases with the 
globule size and decreases with the ionizing flux. At distances from the ionizing star larger than $0.06 - 0.08$\,pc, 
it is $\tau_d^0 \simeq 1 - 2$. This corresponds to the range of values  for $\tau_{R_i}$
mentioned above. In fact, under the assumption that the neutral, inner part of the globule 
has constant density, and as long as $R_0<<R_i$, $\tau_d^0$ coincides with the radial optical depth $\tau_{R_i}$. In this case, Equation~(11) links $\tau_{R_i}$ to the ionizing flux and to the globule radius. Both parameters can be easily obtained, or constrained, by the observations.

\subsection{Radiation field inside the globule}
Assuming the number density of globules present 
within the HII region to be low enough that each one of them
is directly exposed to the radiation coming from the central star, the primary sources of
energy within a globule of radius $R$ will be: 
\begin{enumerate}
\item
The radiation (both ionizing and nonionizing) from the star exciting the HII region:
\begin{equation}
L_g^s (d) = L_s \frac{{R ^2 }}
{{4d^2 }};
%\eqno{(6)}
\end{equation}
\item
The diffuse radiation, including both \Lya\, and other nebular radiation:
\begin{equation}
L_g^{nebular} = \frac{3}
{4}\rho L_s \left( {\frac{{R }}
{{R_{HII} }}} \right)^2 
%\eqno{(7)}
\end{equation}
\item
The radiation $L_\star$ from the star at the center of the globule.
\end{enumerate}
Within the globule, the energy density results from a combination of these three terms, 
plus the thin shell discussed in the previous section, each one attenuated by the dust and gas
absorption. Note that now the first term dominates over the second at distances $d\le R_{HII}/\sqrt{3\rho} 
\simeq 0.9 R_{HII}$. 
We neglect the IR radiation produced by the globule itself as a significant source
of dust heating. Figure~\ref{fluxes_in_the_globule} provides a sketch of the contributions
we are going to consider, beginning with the ionized wind. We shall assume hereafter $R\simeq R_i$.

\subsubsection{Ionized wind}
Figure~\ref{thinshell} illustrates how, according to the model
described in the previous section, the UV flux is absorbed by the gaseous and dusty components 
within the ionized wind. We have assumed $d=0.1$\,pc, corresponding
to $\simeq45$\arcsec\, at the distance of the Orion Nebula and $R_i=100$\,AU. With this choice of parameters,
the FUV radiation is only $\approx 35\%$ absorbed before reaching the ionization front. 
The FUV radiation also traces 
the fraction of EUV radiation absorbed by the dust\footnote{In this paper we assume $Q_\nu=1$ for the dust absorption efficiency in the EUV, FUV, \Lya, and nebular radiation. The corresponding optical depths are therefore equal.}. 
The dust optical depth, depending linearly on the density, dominates the absorption of
EUV photons for distances larger than $r \approx 1.3 R_i$. When the gas optical depth (proportional to $n^2$) 
becomes dominant, the ionizing flux has already been reduced by approximately
one half. The fraction of EUV radiation intercepted by the gas is therefore predominantly absorbed in a 
relatively thin layer outside the ionization front.
%;and re-emitted with most of the energy in 
The layer acts as a secondary
source of recombination radiation, trapped between the neutral PDR region inside and the 
ionized wind outside. Each side receives fifty per cent of the recombination radiation, 
again with \Lya\, carrying $\simeq 1/3$ of the energy and FUV and optical lines carrying the rest. 
%This secondary source is not negligible. 
A crude estimate indicates that the flux emitted in each direction is approximately
$1/2 \times 1/2 \times 1/2 = 1/8$ of the flux directly coming from the ionizing star, the three factors 
accounting for the fraction of the energy emitted 
in the EUV, the fraction of EUV radiation absorbed by the shell, and the fraction of photons emitted up/downstream.
This amount is not negligible and has been included in the model.

We treat the radiative flux within the ionized wind in a 1-d approximation.
The flux, $F_s$, coming from the ionizing star has a cylindrical symmetry along the globule axis, defined by the line joining the 
globule center to the ionizing star and equations~(1) and (2) can be used with 
the optical depth, $\tau_d^\infty$, introduced in Section~3.2. 
The nebular radiation flux, $F_n$, as well as the radiation produced at 
the ionization front, $F_{if}$,
are regarded as uniform, extended sources at the outer and inner edge of the ionized region, respectively.
Treating the dusty region as a plane parallel slab illuminated by two uniform radiation fields, 
we use the attenuation averaged over $2\pi$ steradian:
\begin{equation}
\left\langle e^{-\tau}\right\rangle_{2\pi}  = \int\limits_0^1  e^{-\tau/x} dx
\end{equation}
In general, all nebular \Lya\, photons entering the ionized wind 
will be rapidly absorbed by the dust, whereas a fraction, 
$\left\langle e^{-\tau_d^\infty}\right\rangle_{2\pi}$, of the 
nebular radiation will cross the region. 

To these terms, we add, together with the flux from the central star, $F_\star$,  
the fraction of nebular recombination lines, \Lya\, excluded, coming from the 
back side after crossing the neutral globule. 

\subsubsection{Neutral globule}
Within the neutral region, we use a 2-d treatement. No EUV radiation is present, 
and the fraction of FUV 
radiation crossing the ionization front is further attenuated  by an optical depth, 
\begin{equation}
\tau _{2d} (r,\theta ) = \tau_{R_i} \left[ {1 + \left(\frac{r}{R_i}\right)^2  - 2\left(\frac{r}{R_i}\right)\cos (\theta  - \theta _{h}) } \right]^{1/2}, 
\end{equation} 
between the point $Q(\theta_h)$ at the globule surface and the point $P(r,\theta)$, along the line parallel to
the globule axis. It is $\theta _h  = \arcsin \left[{r\sin\theta/R_i} \right]$ (Figure~\ref{globule_geometry}). Note that since the globule is homogeneous, distances and optical depths
are related by the simple relation:
\begin{equation}
\tau_r=\tau_{R_i} \frac{r}{R_i}.
\end{equation}

On the hemisphere facing the ionizing star, the ionized shell converts UV photons into recombination line photons.
Like the ionized wind, the neutral globule is optically thick to \Lya\, photons and translucent to recombination lines 
at the range of optical depths we are dealing with. 
To illustrate their propagation, we consider first the energy input, $E$,
%~(erg s$^{-1}$Hz$^{-1}$) 
from an isotropic radiation field with specific intensity, 
$I_\nu$,
%~(erg cm$^{-2}$s$^{-1}$sr$^{-1}$Hz$^{-1}$)
on the surface of radius, $R_i$:
\begin{equation}
E_\nu=(4\pi R_i^2) \pi I_\nu.
\end{equation}
Within the globule, the radiation field is attenuated by a factor:
\begin{equation}
\left\langle e^{-\tau^\prime(r)}\right\rangle_{sphere}=\frac{1}{2}\int_0^\pi e^{-\tau^\prime(r,\theta^\prime)}\sin\theta^\prime d\theta^\prime,
\end{equation}
where $\tau^\prime(r,\theta^\prime)$ is the optical depth between the point $R$, assumed for 
simplicity on the globule axis $(\theta=0)$ and 
a point $S(\theta^\prime)$ on the surface (see Figure~\ref{globule_geometry}). 
Figure~\ref{Lyalphainglobule_NEW_2} shows that
when the globule is optically thick, the energy density at the center goes to zero, whereas at the surface it approximates 50\% of its value in the open space, as the flux
comes only from the outer side. Continuity across the globule edge is preserved, as the outer energy density
is also reduced in the globule shadow with respect to the value in open space. The attenuation in the vicinity of the surface can then be estimated integrating Equation~(20) between 0 and $\pi/2$ in the limit $\tau_{R_i}>>\tau$, $\tau$ being the radial optical depth from the surface. In this case, Equation~(20) reduces to Equation~(16).

Since the two hemispheres are illuminated by different sources of radiation, isotropy is broken, 
and Equation~(20) must be replaced by an equation weighting the two contributions separately. The flux at a point
$P$ will now depend also on the angle $\theta$ (see Figure 4). 
It is convenient to write Equation~(19) as
\begin{equation}
E_\nu=\pi a^2 (4\pi I_\nu),
\end{equation}
and split the isotropic flux intercepted by the sphere cross section
in two terms:
\begin{equation}
4\pi I_\nu = \Omega_1 A_1 I_\nu^1 + \Omega_2 A_2 I_\nu^2.
\end{equation}
Here, $\Omega_i$ are the solid angles subtended by the areas radiating with intensity, $I_i$, and $A_i$ are
the corresponding mean attenuations. % Simple geometric relations (APPENDIX?) show that 
It is 
\begin{equation}
A_1(r,\theta)= \left\langle e^{-\tau^\prime(r,\theta)}\right\rangle_{up}=
\frac{1}{4\pi}\left[
\int_0^{2\pi}\int_0^{\theta_{2\pi}^\prime(r,\theta,\phi^\prime)} 
   e^{-\tau^\prime(r,\theta,\theta^\prime,\phi^\prime)}\sin(\theta^\prime)d\theta^\prime d\phi^\prime\right],
\end{equation}
and,
\begin{equation}
A_2(r,\theta)= \left\langle e^{-\tau^\prime(r,\theta)}\right\rangle_{down}=
\frac{1}{4\pi}\left[
\int_0^{2\pi}\int_{\theta_{2\pi}^\prime(r,\theta,\phi^\prime)}^{2\pi} 
   e^{-\tau^\prime(r,\theta,\theta^\prime,\phi^\prime)}\sin(\theta^\prime)d\theta^\prime d\phi^\prime
\right].
\end{equation}
The optical depth, $\tau^\prime(r,\theta,\theta^\prime,\phi^\prime)$, between the point, $P(r,\theta)$, inside 
the globule and a point, $S(\theta^\prime,\phi^\prime)$, on the surface of the sphere is related through Equation~(18) to $r^\prime$, solution of 
the equation 
\begin{equation}
{r^\prime}^2+2r\left(\cos\theta \cos\theta^\prime+\sin\theta \sin\theta^\prime \cos\phi^\prime\right)r^\prime
+r^2-R^2=0,
\end{equation}
whereas the angle, $\theta_{2\pi}^\prime(r,\theta,\phi^\prime)$, at $P$ between  the direction of the ionizing star, 
and the
globule's equator is given by the equation:
\begin{equation}
r^2 \cos^2\theta \tan^2\theta^\prime_{2\pi}-2r^2\sin\theta \cos\theta\cos\phi^\prime\tan\theta^\prime_{2\pi}+r^2\sin^2\theta-R^2=0.
\end{equation}

\subsubsection{Flux from the central star}
The flux from the central star depends on $r^{-2}$ and on the radial extinction. 
Since the central stars of the proplyds in the Orion Nebula typically have late
spectral types, we account for the corresponding reduction of the dust optical depth 
through the ratio of absorption efficiencies $Q(\lambda_{T_\star})/Q_{UV}\cong Q(\lambda_{T_\star})$, where $\lambda_{T_\star}$ is the effective 
wavelength of the photospheric radiation. At distances larger than $R_0$, the stellar flux is
\begin{equation}
F_\star(r) = \frac{L_{\star}}{4\pi r^2}e^{ -\tau_r(r)Q(\lambda _{T_\star})}.
\end{equation}
where $\tau_r$ is the radial optical depth from the central star.
Figure~\ref{Fshell_theta1C} shows the flux densities along the symmetry axis 
for a globule again with $R_i=100$\,AU and $d=0.1$\,pc. For the star, we assumed 
$T_\star=4,000$\,K and $R_\star = 2.5R_\sun$. We shall refer hereafter to these values as the standard case. Within the ionized
wind, the dust heating is dominated by the FUV flux. Inside the globule, the dominant source of heating 
is the radiation from the central star, in the inner parts, and the \Lya\, radiation closer to the ionization front.

\subsection{Infrared emission}
The dust temperature, $T_d(r,\theta)$, is determined by the local energy balance of dust grains.
Within the neutral part of the globule, it is given by the equation
\begin{equation}
\sigma_g 
\left\{ Q_{UV} 
\left[ F_{FUV} (r ,\theta ) + F_{Ly\alpha}(r,\theta)+F_{REC}(r,\theta) \right] 
     + Q(\lambda_{T_\star}) F_\star(r) 
\right\}   = 4\sigma_g \int_0^\infty  {Q(\nu )\pi B\left[\nu ,T_d(r,\theta)\right]d\nu }, 
%\eqno{(10)}
\end{equation}
where $F_{REC}$ refers to the recombination radiation emitted at the ionization front and $\sigma_g=\pi {a_g}^2$ indicates the geometrical cross section of a typical grain of radius $a_g$. A similar equation holding for the ionized 
flow includes the EUV flux, and, in our approximation, has the distance, $r$, as the only independent variable.

The dust temperature depends strongly on the assumed grain parameters.
Following Natta \& Panagia (1976), we may take for reference ``type-1'' grains 
with $a_g=a_1=0.1\,\micron$ and emissivity efficiency in the infrared
\begin{equation}
Q(\nu) = \frac{{2\pi a_1 }}{\lambda },
\end{equation}
and the smaller ``type-2'' grains with $a_g=a_2=0.02\,\micron$ and emissivity efficiency
\begin{equation}
Q(\nu ) = 7 \times 10^{ - 4} \frac{{a_2 }}
{{\lambda ^2 }}.
\end{equation}
Type-1 grains have properties similar to the classical Draine and Lee (1984) silicates.
Power-law emissivities allow analytical solutions to Equation~(28) for $T_d$.
Writing the last two equations as
\begin{equation}
Q(\nu)= Q_0 \lambda ^{-\beta}, 
\end{equation}
the black body averaged dust opacity $Q_\beta$ is
\begin{equation}
Q_\beta T_d^{4+\beta}   = \int_0^\infty  {Q(\nu )\pi B(\nu ,T_d )d\nu  = } 
\left\{ 
\begin{array}{lc}
  1.510 \times 10^{ - 4} Q_0 T_d ^5 \mbox{\hskip 1cm}\beta \mbox{ = 1;} \hfill \\
  5.150 \times 10^{ - 4} Q_0 T_d ^6 \mbox{\hskip 1cm}\beta \mbox{ = 2.} \hfill \\ 
\end{array}  
\right.
%\eqno{(11)}
\end{equation}
Equation~(28) therefore becomes
\begin{equation}
T_d(r ,\theta ) = \left\{   
{\frac{
      \left[ F_{FUV} (r ,\theta ) + F_{Ly\alpha}(r,\theta)+F_{REC}(r,\theta) \right] + Q(\lambda_{T_\star}) F_\star(r) 
      }
     {4 \cdot Q_\beta}
} 
\right\}^{1/(4+\beta)},  
%\eqno{(16)}
\end{equation}
and a similar equation holds for the ionized wind. In Figure~\ref{Tglobule_axis}, we compare the temperature
distributions along the globule axis resulting from the two types of grains, assuming the standard parameters.
The smaller grains reach
temperatures more than twice that of the larger grains, i.e $T_d\simeq 260$\,K
vs. $T_d\simeq 120$\,K. Also, with the exception of the region closer to the
central star ($r\lesssim 1/3 R_i$), the temperature is almost constant due to the weak dependency of the temperature on the flux.
In the rest of this paper, we shall use Type-1 grains to be consistent with our treatment of the disk, as also assumed by  Chiang and Goldreich (1997, CG97).
This choice allows us to use Equation~(33) to obtain the dust temperature and to apply the tabulated dust efficiency of 
Draine and Lee (1984) to calculate a more realistic spectral energy distribution. We remark here
that scattering of stellar radiation from dust grains is not considered in our model.

The emitted IR spectrum is given by the integrals:
\begin{equation}
L_{IR} (\nu ) = \int_{R_0 }^{R_i }\int_0^{\pi } 
\left\langle {e^{ - \tau (\nu )} } \right\rangle 
n_d\pi a_g ^2 Q(\nu )\pi B\left[\nu ,T_d(r,\theta )\right]
2\pi r^2 \sin\theta  d\theta  dr, 
%\eqno{(17)}
\end{equation}
in the neutral part of the globule,  and 
\begin{equation}
L_{IR} (\nu ) = \int_{R_i }^{R_\infty } 
\left\langle {e^{ - \tau (\nu )} } \right\rangle 
n_d(r)\pi a_g ^2 Q(\nu )\pi B\left[\nu ,T_d  (r)\right] \pi R_i^2 dr
%\eqno{(17)}
\end{equation}
for the ionized wind. To keep our treatment as simple as possible, we neglect the 
absorption of IR radiation in the energy budget of the envelope. 
This assumption is valid within  the range of the optical depths in the UV and allows us
to use Equation~(33) avoiding the iterative computation of the IR radiative transfer. 
On the other hand, the 10\,\micron\, silicate feature can be self-absorbed even at our relatively modest optical depths, having an absorption cross section comparable to that at 2.2\,\micron. 
Therefore, we introduced the exponential factors in Equations~(34)-(35) to account for this effect.

Figure~\ref{IR_FLUX_envelopeonly} shows the Spectral Energy Distributions (SEDs),
$\nu F_{IR} (\nu )$, calculated for the Draine and Lee (1984) dust grains and, for comparison, for the
Panagia's type 2 grains. The resulting SEDs peak at $\approx 25$\,\micron and $\approx 10\micron$, respectively.
Depending on the grain properties, the emission can be dominated by neutral globule or by the ionized shell.

\section{Externally illuminated disks}
Our next step is to evaluate the IR SED of circumstellar disks embedded in HII regions.
Although standard models for T-Tauri disks assume the star+disk systems to be isolated, a number of studies have been devoted to the effects of the surrounding environment on the disk thermal structure and emitted spectrum. 
In particular, Natta (1993) and Butner, Natta, \& Evans (1994) modeled the disks' thermal emission when they are surrounded by envelopes.
Both studies find that at large distances from the central star the disks become significantly warmer than if direct heating alone was considered. The heating mostly affects the long wavelength emission from the disk. One may expect to obtain similar results for the disks in an HII region.

As for the globules, the disk temperature depends on a combination of heating by the central star in the disk, external heating due to the
radiation directly coming from the ionizing star, and heating from the nebular environment. We neglect heating from accretion, which should be a minor component outside a few stellar radii.

The radiation from the nebular environment typically
hits the disk surface at angles much larger than $\alpha$,
the grazing angle of the radiation coming from
the central star. 
In particular, the flux coming directly from the ionizing star
depends on the tilt angle, $\theta_d$, of the disk axis with respect to 
the direction of the ionizing source  and acts only on one of the two disk faces. 
The nebular radiation, isotropic and
roughly constant within the nebula, flows uniformly 
into the entire disk surface. 
For a geometrically thin, optically thick disk it is $\alpha \approx 0.4 R_* / a$
(Ruden \& Pollack 1991) and the energy input 
from the disk's central star will be dominant over the 
external radiation at disk radii  
%\begin{equation}
\begin{eqnarray}
a_d  &\leq &R_{\star}  \cdot \left[ {\frac{{L_s }}
{{4\pi d^2 }}\cos (\theta_d ) + \frac{{3\rho L_s }}
{{16\pi R_{HII} ^2 }}} \right]^{-1/3} \left( {\frac{{3\pi }}
{{2\sigma T_\star} ^4 }} \right)^{-1/3} \\ 
     &&=  1.4\left(\frac{R_\star}{R_\odot}\right)
            \left[  \frac{\cos(\theta_d)}{d_{0.1}^2}+0.26 \right]^{-1/3}
            \left(\frac{T_\star}{T_\odot}\right)^{4/3} \hspace{.3cm}{\rm AU}
\end{eqnarray}
%\end{equation}
on the side facing the ionizing star,  or 
%\begin{equation}
\begin{eqnarray}
a_d & \leq &R_{\star}  \cdot \left( {\frac{{3\rho L_s }}
{{4\pi R_{HII} ^2 }}} \right)^{-1/3} \left( {\frac{{3\pi }}
{{2\sigma T_{\star} ^4 }}} \right)^{-1/3} \\
     &&=7.1\left(\frac{R_\star}{R_\odot}\right)
            \left(\frac{T_\star}{T_\odot}\right)^{4/3} \hspace{.3cm}{\rm AU},
\end{eqnarray}
%\end{equation}
on the opposite face, having used the standard parameters and having indicated with 
$d_{0.1}=d$(pc)$/0.1$\,pc. 
In Figure~\ref{Inoutdiskface}, $a_d$ is plotted as a function of the disk tilt angle $\theta_d$ for various
distances from the ionizing star. The dashed line represents the limit of pure
nebular illumination. This occurs a) on the disk face opposite to the ionizing star, b) on both faces if
the disk is oriented edge-on with respect to the ionizing star, ($\theta_d\simeq 90^\circ$),
3) at large distances from the ionizing star.
Figure~\ref{Inoutdiskface} clearly shows that, with our standard set of parameters, the heating of typical 
circumstellar disks is generally dominated by the nebular environment at distances larger than a few AU.

In reality, the EUV flux hardly reaches the 
disk surface in environments like the core of the Orion Nebula, 
since the neutral photoevaporated wind pushes away the ionization front from the disk. 
All radiation fields, including the radiation from the disk's central star, are 
attenuated by the evaporated wind. In the next section, devoted to the actual predictions for 
the SEDs of photoionized globules in the Orion Nebula, this effect will be explicitely taken into account.
Here, however, we shall not reduce the external fluxes by generic absorption coefficients.
This because, on the one hand, the main goal of this section is to illustrate how we build the model, and we want 
to keep the formalism as close as possible to our assumed reference model for circumstellar disks. On the other hand, at large distances from the ionizing star, photoevaporation can be so negligible that the diluted EUV flux can actually reach the disk surface.
Prominent photoevaporated proplyds in the Orion Nebula are clearly concentrated within $\approx30\arcsec$ from \thC.
In particular, one does not expect the rear face of the disks to be photoevaporated, as it is heated only 
by the disk's central star, and by the nebular radiation possibly attenuated by the evaporated material streaming away from the system (the cometary tails of the proplyd). The dark disks seen in silhouette against the nebular background do
not show evidence for photoevaporeted winds (McCaughrean \& Stauffer 1994). In principle, they can can be modeled using the equations presented in this section.                                                                                                                               
The reader is cautioned, in any case, that a comparison with real sources should take into account absorption effects{\footnote{For reference, the envelope discussed in the previous section reduces the flux from the ionizing star by a factor $\approx 1/5$, an amount comparable to moving the source twice as far.}. 

\subsection{Disk Spectral Energy Distributions}

We calculate the disk emission using our modified version of the CG97 model, which provides a self-consistent treatment of the vertical hydrostatic equilibrium and radiative transfer of a disk. Disk flaring results from the hydrostatic equilibrium of the disk material
in the gravitational field of the central star (Shakura \& Sunyaev 1973, Kenyon \& Hartmann 1987).
In the CG97 model, radiation from the central star grazes the disk surface at an angle $\alpha$ and is  
entirely absorbed once it has reached a tangential optical depth $\tau_\parallel\simeq 1$
at the characteristic wavelength of the stellar radiation. This absorption produces an optically thin layer of  ``superheated'' dust grains, since the grain emissivity in the infrared is lower than the absorption efficiency in the visible. Half of the radiation emitted by the layer goes inward and regulates the inner disk temperature.

We have modified this scheme to account for the external radiation field. Since the external radiation usually hits the disk at an angle much larger than $\alpha$, photons entering along the normal to the disk surface will be absorbed deeper in the disk than those coming from the central star. We introduce a perpendicular optical depth $\tau_\bot$ related to the tangential optical depth
by the equation $\tau_\bot\approx\tau_\parallel\alpha$. 
Instead of referring to the $\tau_\parallel=1$ limit for the grazing radiation as CG97, we use the optical depth, $\tau_\bot$ as the variable that governs
the propagation of the radiation through the disk. In this way, 
the dust temperature can be estimated at any disk radius, $a$, as a function of $\tau_\bot$ instead
of the single value estimated at $\tau=1$ by CG97. 

With these assumptions, the flux from the central star is attenuated by a factor,  
$e^{ - \tau_\bot Q(\lambda_{T_\star})/\alpha(a)}$, 
within the disk,  whereas the flux from the ionizing star is reduced by a factor, 
$e^{ - \tau_\bot/\cos\theta_d } $. To be consistent with the notation used in the previous sections, $\tau_\bot$ is now the optical depth at UV wavelengths. In principle, the disk flaring produces a modulation of $\theta_d$, but we ignore this effect considering that the variations average to zero over the circumference. 
We also assume that all the radiation coming from the ionizing star is locally absorbed by the dust, directly or through absorption of recombination radiation.
It is, therefore,
\begin{equation}
F_d^s(\tau_\bot ) = \frac{{L_s }}
{{4\pi d^2 }}\cos (\theta_d ) \cdot e^{ - \tau_\bot/\cos\theta_d }.
\end{equation}

The attenuation of the nebular radiation within the disk
can be readily expressed as a function of $\tau_\bot$ by using Equation~(16), assuming the disk curvature is 
negligible: 
\begin{equation}
F_{nebular} (\tau_\bot ) = \frac{\rho L_s}{8\pi R_{HII}^2}
\left[\int_0^1 {e^{ - q\tau_\bot /x} dx}+2\int_0^1 {e^{ -\tau_\bot /x} dx}\right], 
\end{equation}
where $x$ is the distance from the normal to the disk surface. The two terms on the right side represent the \Lya\, and
the nebular lines, respectively.
The total external flux on the disk face pointing to the ionizing star is therefore
\begin{equation}
F_{out} (\tau_\bot ) = F_d^s(\tau_\bot )+F_{nebular} (\tau_\bot ),
\end{equation}
whereas on the other face only the term containing $F_{nebular} (\tau_\bot )$ will be present.

The external radiation affects the geometry of the disk. In particular, the grazing angle, $\alpha(a)$, 
depends on the external flux. Following CG97, it is:
\begin{equation}
\alpha (a) \approx \frac{{\alpha _1 }}
{a} + a\frac{d}
{{da}}\left( {\frac{H}
{a}} \right),
%\eqno{(19)}
\end{equation}
where $\alpha_1\simeq0.4R_\star$, and $H$ is the disk height above the midplane. 
For a disk vertically isothermal, it is
\begin{equation}
\frac{H}
{a} \approx 4\left( {\frac{a}
{{R_\star  }}} \right)^{2/7} \left( {\frac{{T_e }}
{{T_c }}} \right)^{4/7}.,
\end{equation}
where the effective temperature, $T_e$, is a measure of the energy input at the disk surface and $T_c=GM_\star\mu/(kR_\star$), $\mu$ being the mean molecular weight of the gas. 
%In absence of an outer radiation field,  it is:   
%\begin{equation}
%T_e(a)  \approx \left( {\frac{\alpha }
%{2}} \right)^{1/4} \left( {\frac{{R_\star  }}
%{a}} \right)^{1/2} T_\star,  
%\end{equation}
%so that 
%\begin{equation}
%\alpha (a) \approx \frac{{\alpha _1 }}
%{a} + \alpha _2 a^{2/7}, 
%\end{equation}
%with 
%\begin{equation}
%\alpha _2  =  \frac{8}{7}  \left({\frac{{T_\star  }}
%{{T_c }}} \right)^{4/7} R_\star  ^{ - 2/7}. 
%\end{equation}
%where $T_c=GM_\star\mu/(kR_\star$).	
If an outer radiation field is present, it is 
\begin{equation}
T_e(a)  \approx \left[ {\left( {\frac{\alpha }
{2}} \right)\left( {\frac{{R_\star  }}
{a}} \right)^2 T_\star ^2  + \frac{{F_{out} (0)}}
{\sigma }} \right]^{1/4} .
\end{equation}
With this expression for $T_e$, Equation (43) cannot be solved analytically. In the limit of outer radiation field dominating the disk heating (i.e. at large distances from the disk central star), 
the first term can be neglected and we have:
\begin{equation}
\frac{H}{a} \approx 4\left( \frac{a}{R_\star} \right)^{1/2} 
\left\{ \frac{\left[F_{out}(0) /\sigma\right]^{1/4}}{T_c} \right\}^{1/2}. 
\end{equation}
In this case,
\begin{equation}
\alpha (a) \approx \frac{{\alpha _1 }}
{a} + \alpha _3 a^{1/2} 
%\eqno{(20)}
\end{equation}
with
\begin{equation}
\alpha _3  = 2\sqrt {\left[ {\frac{F_{out}(0)}
{\sigma }} \right]^{1/4} \frac{1}
{T_cR_\star} }. 
\end{equation}
In Figure~\ref{alpha}, we show the vertical disk profiles calculated for a disk facing the ionizing star at $d=0.1$\,pc, with no extinction in front of the disk surface. In this situation, the flaring calculated on both faces, i.e. with and without the EUV radiation, is significantly higher than that resulting from the standard CG97 model. In practice, these disk profiles are
only indicative of an optical depth surface, since when photoevaporation is present there is no discontinuity between the disk surface and the envelope. A large flaring angle can rather be  regarded as an indication for disk photoevaporation. The case with no direct radiation 
from the ionizing star provides the minimum flaring profile for the back surface of a disk within the HII region (assuming negligible extinction from the envelope material).

The disk flaring will be dominated by the external flux at distances larger than
\begin{equation}
R_{flaring}  = \left( {\frac{4}{7}} \right)^{\frac{{14}}{3}} 
               T_{\star}^{\frac{8}{3}} T_c ^{\frac{-1}{3}} 
               \left( {\frac{F_{out}}{\sigma } } \right)^{-7/12} R_{\star}. 
\end{equation}
With our typical parameters, $R_{flaring}  \simeq 2\mbox{AU}$. 

The temperature of the superheated dust layer is given by 
the combination of internal and external contributions  at 
the optical depth, $\tau_\bot$:
\begin{equation}
T_{ds} (\tau_\bot ,a) = \left[ {
\left( {\frac{{R_{*}}}{a}} \right)^2 \sigma {\rm T}_{*} ^4  
e^{- \tau_\bot Q(\lambda_{T_\star)}/\alpha (a)}
+ \frac{2}{\alpha (a)} F_{out}(\tau_\bot )
                           } \right]^{\frac{1}
{{4 + \beta}}} \left( {\frac{{T_{d1} ^\beta}}
{{2\sigma }}} \right)^{\frac{1}
{{4 + \beta}}} ,
%\eqno{(21)}
\end{equation}
where
\begin{equation}
T_{d1}  = \frac{{hc}}{{8\pi a_g k}}
\end{equation}
is a characteristic temperature of the dust grains, related to the black body averaged dust emissivity 
%$\varepsilon(\lambda ) = (2\pi a_g /\lambda )^\beta  $
at temperature $T$ by 
\begin{equation}
\bar Q(T)  = \left( {\frac{T}
{{T_{d1} }}} \right)^\beta.  
\end{equation}

In Figure~\ref{Tds_NOENVELOPE}, we plot the temperature distribution calculated at various distances from the central
star as a function of the vertical optical depth. By comparison, the CG97 approximation of uniform temperature across the superheated layer:
\begin{equation}
T_{ds}  = \left[ {\left( {\frac{{R_{\star} }}
{a}} \right)^2 \sigma T_{\star} ^2  + F_{out} (0)} \right]^{1/(4 + \beta )} 
\left( {\frac{T_d1}{4\sigma} }\right)^{1/(4 + \beta)},
\end{equation}
would have provided 
$T_{ds}=355$\,K, 220\,K, 147\,K, and 113\,K for the $a=3, 10, 30, 100$\,AU, respectively.

The flux emitted by the superheated layer will be given by  
\begin{equation}
F(a,\nu ) = \int_0^3 {e^{ - \tau_\bot^\prime \cdot Q(\nu)  } Q(\nu)  B_\nu \left[{\nu,T_{ds}(\tau_\bot^\prime,a)}\right]
d\tau_\bot^\prime} 
\end{equation}
having assumed that all the flux is absorbed when $\tau_\bot=3$. The luminosity density from the superheated layer, normalized to the stellar luminosity $L_\star$, is
\begin{equation}
L_{ds}(\nu)  = \frac{8\pi ^2 \nu }{L_\star}\frac{1}{2} \int_{a_i }^{a_o } {F(a,\nu )da},
\end{equation}
where $a_i=0.07$\,AU and $a_o=100$\,AU are our assumed inner and outer disk radii, and we have taken into account that only
1/2 of the flux is radiated in the outer space.

In what concerns the inner disk, 
we adopt the same scheme of CG97 with modifications needed to account for the ambient radiation. Equations 12a-c of GC97 now become:
\begin{equation}
%\begin{mathletters}
%\begin{eqnarray}
\begin{array}{lr}
  T_{in}(a)  \approx \frac{1}
{{2^{1/4} }}\left[ {c_{4} ^4 a^{ - 3/2}  + \frac{{F_{out} (0)}}
{\sigma }} \right]^{1/4} \hfill \mbox{region $a$} \\
  T_{in}(a)  \approx \frac{{T_{d1} ^{\beta /(4 + \beta )} }}
{{\left( {2K_V \Sigma _0 } \right)^{1/(4 + \beta )} }}\left[ {c_{4} ^4  + \frac{{F_{out} (0)}}
{\sigma }a^{3/2} } \right]^{1/(4 + \beta )} \hfill \mbox{region $b$ } \\
  T_{in}(a)  \approx \frac{1}
{{2^{1/(4 + \beta )} }}\left[ {c_{4} ^4 a^{ - 3/2}  + \frac{{F_{out} (0)}}
{\sigma }} \right]\left[ {\left( {\frac{{R_ *  }}
{a}} \right)^2 \sigma T_ *  ^4  + \frac{{F_{out} (0)}}
{\sigma }} \right]^{1/(4 + \beta )} T_{ds} ^{1/(4 + \beta )} \hfill \mbox{region $c$} \\ 
\end{array}
%\end{eqnarray}
%\end{mathletters}
\end{equation}
having defined $c_{4}$ as
\begin{equation}
c_{4}  = \left( {\frac{{\alpha _3 }}
{4}} \right)^{1/4} R_ *  ^{1/2} T_ *.  
\end{equation}
The three regions are limited by the radii $a_{a\vert b}$  and  $a_{b\vert c}$, 
derived by setting $\tau_\bot  \cong 2$ and $\tau_\bot \cong 1.5$, respectively. 
With CG97, we have defined 
\begin{equation}
\tau(a) = \bar Q_{1,2} \kappa _V \Sigma _0 a^{ - 3/2} 
\end{equation}
with $\bar Q_1  = \left( {\frac{{T_{int} }}
{{T_{d1} }}} \right)^\beta  $  and $\bar Q_2  = \left( {\frac{{T_{ds} }}
{{T_{d1} }}} \right)^\beta  $
 for the two conditions,  $\kappa_V  = 400\,\mbox{cm}^{2}~\mbox{g}^{- 1} $, 
and  $\Sigma _0  = 10^3\,\mbox{g\,cm}^{-2} $. Consistent with CG97, we used Equation~(53) for the temperature, $T_{ds}$. 
With our standard parameters, $a_{a\vert b}=140$\,AU. 
The temperature of the internal disk is constant at $T_{d1}\simeq 60$\,K over most of the disk. 
The luminosity density from the disk interior normalized to the stellar luminosity $L_\star$ is now
\begin{equation}
L_{in}(\nu)  = \frac{{8\pi ^2 \nu }}
{{L_{\star} }}\int_{a_i }^{a_o } \left[ {1 - e^{ - \tau_\bot (x,\nu )} } \right]
B\left[\nu ,T_{in}(x)\right] xdx. 
%\eqno{(27)}
\end{equation}

Figure~\ref{Ldisk_NOENVELOPE_front} shows the SED for the front surface of the flared disk. The distribution reaches 
a minimum at $\approx 4$\,\micron, and then rises up to a maximum at $\approx 60$\,\micron. The silicate peaks prominent at 10 and 20\,\micron,
are emitted in the superheated layer.
In Figure~\ref{Ldisk_NOENVELOPE_back}, we present the SED for the face
of the disk opposite to the ionizing star. There is a general reduction of the disk luminosity, both from the superheated atmosphere at short wavelengths and from the interior in the far-IR. 
As we already mentioned, this
type of SED can be representative of the dark silhouettes seen in projection against the nebular background in Orion. For these disks, \Lya\, radiation may play a role only if the disks are still embedded within the HII region, whereas the other nebular radiation is relevant also if the disk is in the PDR region behind the ionization front. One must remember that the Orion nebula is essentially density bounded in the direction of the Earth, being limited by a nonuniform veil of foreground material which is not completely optically thick (O'Dell \& Yusuf-Zadeh, 2000). Dark silhouette disks can therefore be subject to a broad range of radiative input, as shown for example by the presence of [OI] emission in only one object reported by Bally et al. (2000).

\section{Photo-ionized proplyds}
%Disk + Globule}

In this section, we model the thermal emission produced by a photoevaporated disk enshrouded in a dusty photoevaporated envelope. 
We treat the envelope as a dusty photoionized globule of the type discussed in Section 3, whereas the disk treatment will follow the model presented in the previous section. 
However, when the globule is merged with the disk, some of the assumptions adopted for the isolated cases are no longer valid:
\begin{enumerate} 
\item	
The disk is now directly exposed to only a fraction of the outer radiation, due to the globule attenuation. In particular, no EUV radiation reaches the surface of the disk.
\item
The disk is exposed to the radiation reemitted by the globule, mostly in the infrared.
We shall assume that the IR radiation crosses undisturbed the superheated layer and is absorbed in the inner disk.
\item
The feedback contribution of the disk emission to the heating of the globule can be neglected, 
since a passive disk emits mostly in the far-IR, where the dust opacity is negligible.
\item
The radiation emitted by the disk and by the central star is attenuated crossing the envelope. We also take into account the attenuation of the envelope radiation within the envelope itself.
\end{enumerate}

We shall assume the globule to be spherical with the disk lying in the equatorial plane. 
The angles, $\theta_d$, between the disk axis and the direction of the ionizing star, and, $\theta_\earth$, between the disk axis and
the Earth are both critical to model SEDs matching the observational data. They have been taken
into account in an approximate way: the former modulates the amount of stellar radiation falling on the disk surface 
through Equation~(40) but not on the envelope. The latter affects the tilt angle of the disk with respect to the observer.
For simplicity, however, the plots presented in this section have been obtained assuming both $\theta_d=0^\circ$ and $\theta_\earth=0^\circ$.
A disk oriented face-on both to the ionizing star and to the Earth is located along the line of sight of, and behind, the ionizing star.

The globule structure is described by the same set of equations discussed in Section~(3). The only change is 
that the term containing $I_\nu^2$ in Equation~(22) is now equal to zero,
 since the disk, optically thick to ultraviolet radiation, shields the hemisphere
opposite to the ionizing star. In what concerns the radiation flowing into 
the disk, both the fluxes coming from the HII region and from the central star
are attenuated by the globule. In agreement with our previous assumption of FUV dominated flows,
the EUV flux is entirely absorbed before reaching the surface of the disk.  
The radiation propagates
through the superheated layer and then through the disk interior. Only the IR flux re-emitted
by the globule crosses the thin layer undisturbed.

The total IR flux emitted by the globule and reaching the disk surface is 
given by
\begin{equation}
F(a)=\int_0^{2\pi} \int_0^{\theta_{max}(a)}\int_{R_i}^{R_\infty} \int_0^\infty \frac{F_{env}(r,\theta,\phi,\nu)}{4\pi d_a^2}\cos\theta_a r \sin\theta d\nu dr d\theta d\phi
\end{equation}
where 
$F_{env}=n_d\pi a_g^2(\nu)B(\nu,T_d)$ is the flux emitted by a cubic centimeter of dust 
within the globule (cfr. Equations  (34-35)), and  $d_a$ is the distance between the point, $P(r,\theta,\phi)$, within the envelope and a point, $A$, on the surface
of the disk at a projected (equatorial) distance, $a$, form the star and on the plane defined by $\phi=0$. It is 
\begin{equation}
d_a=\sqrt{r^2+\left(2 a \sin\theta \cos\phi -\cos\theta \tan\alpha\right)r+a^2\left(1+\tan\alpha)\right) },
\end{equation}
and $\theta_a$ is the angle between the line, $PA$, and the normal to the disk surface, given by:
\begin{equation}
\theta_a=\sin\alpha \frac{r\sin\theta \cos\phi}{d_a} +\cos\alpha \frac{r \cos\theta - a \tan\alpha}{d_a}
\end{equation}
An accurate estimate of $d_a$ would require an iterative solution of Equation~(60), since the actual position 
of the point, $A$, within the globule depends on the flaring angle, $\alpha(a)$, which in turn 
depends on the radiative heating on the disk. To avoid the rather heavy numerical computation, we adopted the approximate assumption that the position of point, $A$, is
initially given by Equation~(47), this time with the absorption through the envelope explicetely taken into account, and then used the value of $F(a)$ resulting from Equation~(60)
to recalculate the disk flaring angle.

The flux emitted by the disk is calculated following the treatment presented in Section~4 with a few changes. We assume
that the temperature of the superheated layer does not depend on the flux emitted by the globule, as the IR emission propagates almost undisturbed through the disk atmosphere. 
Due to presence of the absorbing envelope, the flaring angle now depends on the distance, $a$,
also via the external flux, $F_{out}$, (Equation 42). 
In general, $R_{flaring}$ (Equation 49) remains of the order of a few AUs over a broad range of parameters, and even when the extinction by the wind/envelope is included, 
as it affects both the flux from the disk central star and from the outer space. 
Within the Orion Nebula the disk flaring is usually dominated by the external flux, so we
shall adopt the approximation of Equation (47).
The luminosity of the superheated layer is given by Equations~(54)-(55). 
In what concerns the temperature of the disk interior, it will depend on the sum of the IR radiation 
received from the superheated layer and from the globule, plus the flux emitted by the disk interior 
itself when the disk is optically thick. Equations~(56)-(59) apply also in this case.

Neglecting the interstellar contribution, the optical depth between $P$ and an observer on Earth is
\begin{equation}
\tau_\nu(r,\phi,\theta,\theta_{\earth}) = \tau_{R_i} Q(\nu)
\left[
(\frac{r}{R_i}\sin\theta \cos\phi + \sin\theta_R \cos\phi_R )^2+(\cos\theta_R-\frac{r}{R_i} \cos\theta)^2 
\right].
\end{equation}
The angles, $\theta_R$, and, $\phi_R$, define the point where the line from $P$ with angle, $\theta_\earth$, intercepts the globule surface. They are related to the coordinates of $P$ and to $\theta_\earth$ by equations:
\begin{equation}
r\sin\theta\sin\phi=R_i\sin\theta_R\sin(\pi-\phi_R)
\end{equation}
and
\begin{equation}
\tan\theta_\earth=\frac{r\sin\theta\cos\phi+R_i\sin\theta_R\cos(\pi-\phi_R)}{R_i\cos\theta_R-r\cos\theta}
\end{equation}
solved numerically. Crossing the ionized wind, the radiation is further attenuated by an optical depth $\simeq Q_\nu\tau_d^\infty$. We neglect the self absorption of the IR radiation emitted by the ionized wind

Figure~\ref{Ldisk+envelope_ALL} shows the SED distribution arising from the front 
face of the disk. With our standard set of parameters, the emission is dominated 
by the disk atmosphere at wavelengths $\lesssim 40\,\micron$~ and by the internal 
disk emission at longer wavelengths. The envelope emission never produces a noticeable 
contribution. The disk atmosphere has a pronounced peak of emission at 3\,\micron\,
due to secondary heating by the envelope. The relevance of this effect depends on the
amount of dust located in the immediate surroundings of the star. We have assumed
uniform density within the globule up to the dust evaporation radius, which is 
of the order of a few stellar radii. A larger dust free cavity would significantly
reduce the intensity of the 3\,\micron\, bump. 

\section{Discussion}
The geometry we have assumed to illustrate the physical basis of our model 
represents in practice an extreme case. A disk oriented face-on with respect to the ionizing star 
receives the maximum amount of radiation. This 
causes a substantial increase of the flaring angle, and therefore of the inner stellar flux intercepted by the disk. 
If the disk is also
oriented face-on with respect to the Earth, the observer
will receive the highest amount of disk radiation. Even if chances of having such a fortunate alignment are rather low, one of the most intriguing 10\,\micron\, sources of the Orion Nebula, SC3 (McCaughrean 
\& Gezari 1991, Hayward et al. 1994; Figure~\ref{th1C}) can be readily interpreted
in this way. With a 10\,\micron\, flux of~$\simeq 4$\,Jy, 
SC3 is the brightest compact source in the Orion Nebula, excepting the BN/KL region. It is located at a projected distance
$\approx 1\arcsec$\,W of \thC, and mostly for this reason it has never been observed at wavelengths shorter than 2\,\micron.
In our 10\,\micron\, images, SC3 is spatially resolved with FWHM $\simeq 0.5\arcsec$, corresponding to approximately
200\,AU, the typical size of circumstellar disks in Orion. 
A direct comparison with the nearby complex of arc-like structures produced by the wind-wind interactions suggests that SC3 is a different kind of phenomenon. Despite the fact that it has the shortest projected distance from \thC, SC3 is apparently undisturbed by the powerful wind from the O6.5 star, as it is 
the only extended source with circular shape. This indicates that the physical 
distance between SC3 
and \thC\, is much larger than for the other sources. The simplest interpretation is that SC3 is a star+disk system in the background with respect to \thC\, with a very high 
mid-IR brightness produced by the disk being oriented face-on  both to \thC\, and to us. Figure~\ref{SEDcompare_distance} shows that the 10\,\micron\, photometry of SC3 is compatible with our standard model and a distance slightly lower than 30"=0.065\,pc. It must also be noticed
that the low resolution 10\,\micron\, spectrum of SC3 obtained by Hayward et al. (1994) shows silicate emission, whereas our standard model predicts at these distances an almost featurless 
SED dominated by the inner disk emission. Further data on this sources are needed to constrain our set of parameters.

In general, disks are randomly tilted with respect both to the ionizing star and to the Earth, and therefore their apparent brightness will be reduced in comparison to our
standard case. In Section~4.1 we stated that the tilt angle with respect to the ionizing star, $\theta_d$, can be simply accounted for by a projection factor $\cos\theta_d$ in the flux received by the disk surface. However, this is strictly true only for tilt angles
$\theta_d<\alpha(a_o)$. Since disks are intrinsically flared, large tilt angles may cause
part of the surface exposed to the ionizing star to go under the edge's shadow. When this happens, the heating of the surface is no longer uniform, the flaring angle varies with the azimuthal angle and the disk loses
its symmetry around the rotation axis. Along its orbit
around the central star, the disk material will experience a periodic modulation of pressure and temperature, and the resulting differential drag may cause the disk to warp. Photoevaporation will occur only 
from the area directly exposed to the flux from the ionizing star, as well as on the opposite disk edge. Photoevaporation from the
edge appears compatible with the observations showing that disks in the Orion Nebula are truncated (McCaughrean, Stapelfeldt \& Close 2000). 
As the analysis of the structure and evolution of circumstellar disks in these conditions goes beyond the scope of this paper, one can maintain the ``$\cos\theta_d$'' treatment to account for a reduced flux falling on the disk surface, 
keeping in mind that the 
derived tilt angle in this case is an overestimate of the real one.

In what concerns the tilt angle with respect to the Earth, $\theta_\Earth$, Chiang and Goldreich (1999) have studied the 
corresponding variation of the SEDs  for isolated disks. They found that the SED shows negligible variations 
until $\theta_\Earth$ approaches the flaring angle at the disk outer radius $\alpha(a_0)$. When  
the inner disk regions enter in the
shadow, the short wavelength emission begins to disappear from the vantage point of the observer. 
To account for this effect, we simply multiply the flux emitted by the disk at each radius by
the fraction of visible area. Figure~\ref{tiltangle} shows the dramatic
reduction of the disk brightness at short wavelengths, together with the stellar radiation, when $\theta_\Earth$ 
increases from $45^\circ$ to 50$^\circ$ crossing $\alpha(a_0)\simeq 48^\circ$.
Our geometrical treatment is only 
valid as long as the disk is optically thick at all wavelengths and large enough to contain a strongly flared surface. 
In the scheme of CG97, the outer disk regions are in a completely optically thin regime (region c of Equation 56),
and therefore the transition is modulated by the variation of optical depth across the disk. 
In our case, however, the truncated disk remains
optically thick (region a) up to the disk edges. We find that the disk remains optically thick up to $a_o \simeq 100$\,AU over a wide range of parameters. In particular, our standard 
disk+envelope model has $a_{a\vert b}=113$\,AU. Thus, the change of SED at short wavelengths is 
much more abrupt. The decrease of disk emission at $\lambda > 30$\,\micron,
is correct as long as the outer parts of the disk are also optically thick and, therefore, 
sensitive to the disk orientation.
In conclusion, the disk orientation with respect to the Earth can be regarded like a ``switch'' for the short wavelength
radiation. If the star is directly visible (e.g. with HST), then the brightness at near- and mid-IR wavelengths
will be high, otherwise if the star is invisible. Transition objects with heavily 
reddened stars should be rare.
Observations at 10\,\micron\, of dark silhouette disks appear to confirm this result: the 
the large dark silhouette 114-426, seen edge-on, has a star under $A_v \ge 60$ magnitudes of extinction (McCaughrean et al 1998) and remained undetected at $[N]\simeq 9$ in our deep survey performed with the UKIRT 3.8\,m telescope (Robberto et al, 2002), whereas the dark silhouette 
with the brightest central star, object 218-354, has been detected at 
10\,\micron\, both by Hayward \& McCaughrean (1997) and by us.

In the standard case, the envelope contribution is negligible  
at all wavelengths with respect to that coming from the disk. 
For some disk orientations, however, the envelope emission becomes important.
This because the envelopes are extended, and the larger the value of $\theta_\Earth$, the 
more the disk emission will be suppressed, whereas the fraction of envelope obscured by the disk will be negligible. 
If the disk is seen close to edge-on, the envelope emission may eventually dominate.
Figure~\ref{4proplyds} shows two of the most prominent objects,
sources 177-341 (HST~1) and 182-413 (HST~10), as observed by with MAX on UKIRT
and, for comparison, with the HST. These systems, with the disks seen edge-on,
are clearly resolved at 10\,\micron\, with some evidence of a central peak. Size and 
orientation in the mid-IR correspond to that observed in the recombination lines. 
These images provide conclusive evidence that photo-evaporated dust can be heated 
in the photoionized globule and emit significantly in the mid-IR, with brightness
comparable to that of the disk. 
It is clear that all possible combinations of angles $\theta_d$ and $\theta_\earth$ may generate a variety 
of SEDs similar to those presented in Figure~\ref{tiltangle}. All are characterized by an SED peaking at $\simeq30-60$\,\micron\, and relatively low fluxes at wavelengths shorter than 
10\,\micron. 

Let us consider the impact of other basic parameters on the SEDs. 
In Figure~\ref{SEDcompare_distance}, we present the SED obtained by placing our standard proplyd 
at various distances between $30\arcsec$ (d=0.065\,pc) and $150\arcsec$ (d=0.33\,pc) 
from the ionizing star. The SEDs are renormalized to the total stellar luminosity to compare their structure 
independently of the geometrical dilution of the signal with the distance. Still, 
the (relative) stellar flux in the visible increases with the distance, due to the corresponding reduction of optical depth 
within the globule described in Section 3.2 (see also Figure~1). In the far-IR the flux drops with the distance because the outer disk is colder, as it receives less flux from the ionizing star, and it is 
therefore flatter, receiving less flux also from the central star.
In comparison, the flux from the superheated layer is less affected by the distance from the ioninzing star,
and for this reason the silicate peak in emission becomes relatively more prominent. 

Figure~\ref{SEDcompare_Rout} shows the variation of the SED with $R_{out}$, the distance of the ionization front. We have considered $R_{out}=100$\,AU, 200\,AU and 300\,AU. The emerging stellar flux is reduced due to the increase of optical depth in the envelope with the
globule size (see again Figure 1). The relative importance of the envelope emission increases significantly, but the 
SED is still dominated by the disk emission due to the special alignment of our standard case mentioned above. To facilitate the comparison with our standard 
case, we have neglected here the contribution from the hemisphere opposite to the ionizing star.

The change of SED with the inner disk radius, $a_i$, is shown in Figure~\ref{SEDcompare_ai}. The gap in the inner disk produces the expected 
decrease of the SED between 2 and 10\,\micron. 
Finally, the variation of SEDs with the stellar temperature is illustrated in 
Figure~\ref{SEDcompare_Tstar}. Since we have kept the stellar radius constant,
this plot also illustrates the variation of SEDs with the luminosity of the central 
star. Reducing the stellar temperature the disk emission becomes more and more important
because the external contribution has been assumed to remain constant.

A direct application of our model is presented in Figure~\ref{datacomparison},
showing a fit to the available photometric data of source 177-341=HST1 
(see Figure~\ref{4proplyds}).
We have assumed $\theta_d=45^\circ$ and $\theta_\earth=75^\circ$, to account for a disk seen nearly edge-on. We used a distance
$d=43\farcs2$, twice the projected distance from \thC.
Hillenbrand (1997) gives a spectral type later than $K6$. We have adopted a $M1$ spectral type with $T_\star= 3750$\,K and radius $R_\star=2.18$\,R$_\odot$. Concerning the geometry, we have assumed
an ionization front at $R_{out}=300$\,AU and a disk with an inner radius $a_i=0.3$\,AU.
To match the stellar photometry, we assumed $A_v=2.2$. The model predicts $\tau_R$=1.45, so that approximately 0.6 magnitudes of visual extinction must be attributed to foreground
extinction. In this fit, the envelope and disk have comparable brightness, the disk emission been provided by the disk interior. 

Given the number of parameters left unconstrained by the available observational data, it is not surprising that with our model we are able to fit a single photometric point at 10\,\micron.  On the other hand, the 10\,\micron\, photometry cannot be readily explained with conventional disk models. Our model also provides a strong prediction for high mid- and far-IR fluxes from these systems. Future high spatial resolution observations at these wavelengths will asses the validity of our assumptions and provide insights into the evolution of circumstellar disks in the most typical environmental conditions.

A number of assumptions deserve some additional comment.
First, we have assumed that the disk atmosphere remains well defined and treatable by the CG97 approximation even
when the flaring angles become extreme (up to $\simeq 45^\circ$). In presence of photoevaporation, however, the 
conditions of hydrostatic equilibrium used to set the disk profile are no longer valid, and the locus of the disk ``atmosphere'' should be recalculated taking into account the radially diverging flow. 
Second, although we have a basic understanding of the globule physics, it is still very hard
to constrain the actual structure from the observations, deriving information such as the density distribution. Third, the grain properties are critical, not only in what 
concerns the absorption efficiencies (Section 3.4), but also the albedo, that we neglected. Scattering processes affect the near- and mid-IR emission of disks
seen nearly edge on. Also, it has been shown that
dust scattering in an optically thin envelope can contribute to the increase of disk temperature at large distances from the central star (Natta 1993), therefore adding a further contribution to the energy input at the origin of the disk flaring.
Finally, we assumed in Section 2 a simplified model for the Orion Nebula.
It is known (O'Dell 2001a, O'Dell 2001b) that the density in the Orion Nebula decreases with the distance from the interface
with the Orion Molecular Cloud, where the main ionization front is located. Also, \thC\, is
approximately three times closer to the main ionization front than to the foreground veil of neutral material, 
where a secondary ionization front is also located. The major effect of a more refined description 
would be a reduction of our estimates for the \Lya\, energy density. We have seen that
this is a secondary source of dust heating. A modulation over the
$r^{-2}$ dependency of the central radiation field would also be introduced. This effect
may be significant in future studies aimed to the detailed comparison of the SED for different objects.
 
\section{Summary and conclusion}
We have explored the IR emission of circumstellar disks in the environment where star formation most typically occurs, i.e. a HII region powered by massive OB stars. This scenario applies in particular to the Orion Nebula, where the interaction of 
circumstellar disk with the environment has  been directly resolved by the HST. 
We have build our model considering the four types of radiation relevant to the dust heating in a HII region, e.g ionizing (EUV) and non-ionizing (FUV) flux from the exciting star (\thC), resonant \Lya\, radiation and the remaining nebular radiation. 
We evaluated how these radiation sources affect the IR SED arising from:
\begin{enumerate}
\item
A spherical, homogeneous, optically thin, circumstellar globule, 
photoevaporated by the UV radiation of the HII region exciting star. The globule has a neutral core of uniform density, and the photoevaporated atmosphere is treated following the Dyson (1968) model.  With our assumptions, the most important parameter constraining
the IR emission, the radial optical depth to the UV radiation, depends only on the distance from the ionizing star and on the globule size.
The thermal emission peaks in the range $10 - 30$\,\micron, depending mostly on the dust composition.
\item
A non-accreting disk directly exposed to the nebular radiation. The disk is in hydrostatic
and radiative equilibrium, and treated following the prescription of Chiang and Goldreich (1997). We modify the CG97 scheme to account for the fact that the nebular radiation hits the disk surface with large angles. We follow the propagation of the various radiative fluxes  through the disk superheated atmosphere, deriving the vertical temperature profile of the disk atmosphere. The disk faces receive unequal amount of radiation and
present different flaring angles. 
In particular, the flux emitted from the face opposite to the ionizing star provides a model to the IR emission of the dark silhouette disks observed in the Orion Nebula.
\item
A combined system composed by the disk and the photoevaporated envelope, i.e. a photoionized proplyd of the type observed in the immediate surroundings of  \thC\, in the Orion Nebula. 
\end{enumerate}
Depending on the distance and on the tilt angle of the disk with respect to the ionizing star, the disk flaring may be substantially higher than in the case of isolated disks. 
The tilt angle with respect to the Earth plays also a major role by hiding the central parts of the disk. The relative intensity of the disk vs. envelope emission
varies with the tilt angle to the direction of the Earth.
The high temperatures reached by the dust either at the disk atmosphere or within the 
envelope produce a SED peaking at $30-60$\,\micron. 

We explore the dependency of the SEDs upon the tilt angle
with respect to the Earth, the distance from the ionizing star, the size on the envelope, the inner
disk radius, and the temperature of the disk's central star.
The resulting SEDs
are characterized by a broad peak of emission at 30-60\,\micron\, and are
in general significantly different from those of isolated disks 
in low-mass star forming regions like Taurus-Auriga.  
Our model indicates that in the presence of 
an external radiation field, relatively evolved ``Class 2" objects 
may display a SED peaking at mid-IR and far-IR wavelengths.
The model explains the strong mid-IR excess we have recently detected on
several sources in a 10\,\micron\, survey of the Orion Nebula. 
 
\section{Acknowledgments} 
The authors whish to thank Bob O'Dell and an anonymous referee for carefull reading of the origial manuscript and useful comments.

\clearpage

\newpage
\begin{figure}	%1
%\figurenum{TEXT}
%\epsscale{NUM}
\plotone{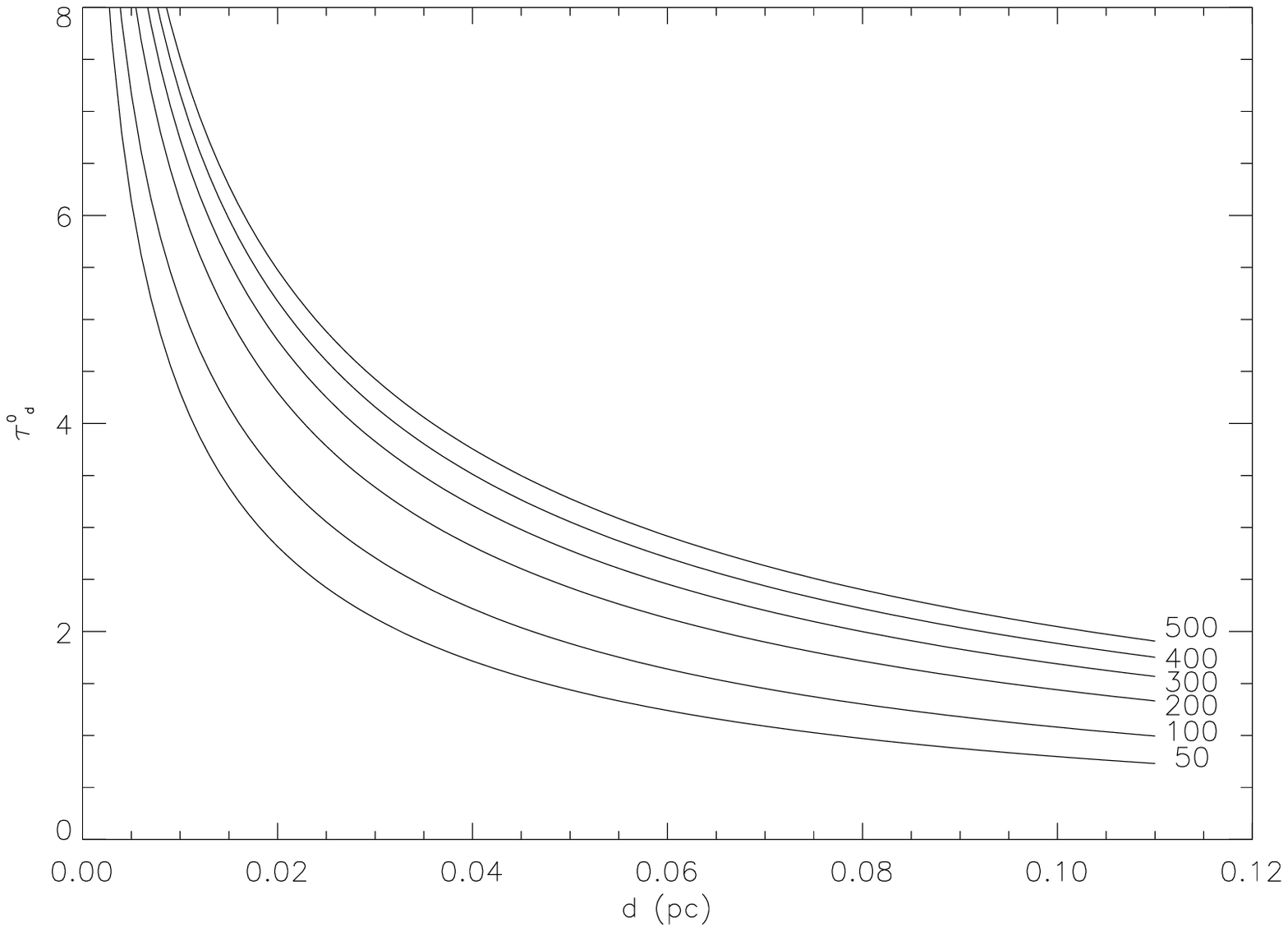}
%\plotone{./Pastor91EQ13_v3.eps}
%\plottwo{EPSFILE}{EPSFILE}
%\plotfiddle{EPSFILE}{VSIZE}{ROT}{HSF}{VSF}{HTRANS}{VTRANS}
\caption{Optical depth $\tau_d^0=\sigma_dn_iR_i$ as a function of the 
distance $d$ from \thC. Six different globules are considered, with ionization
front at $R = 50$, 100, 200, 300, 400, and 500 AU.\label{Pastor91EQ13_v3}}
\end{figure}

\newpage
\begin{figure} %2
%\figurenum{TEXT}
%\epsscale{NUM}
\plotone{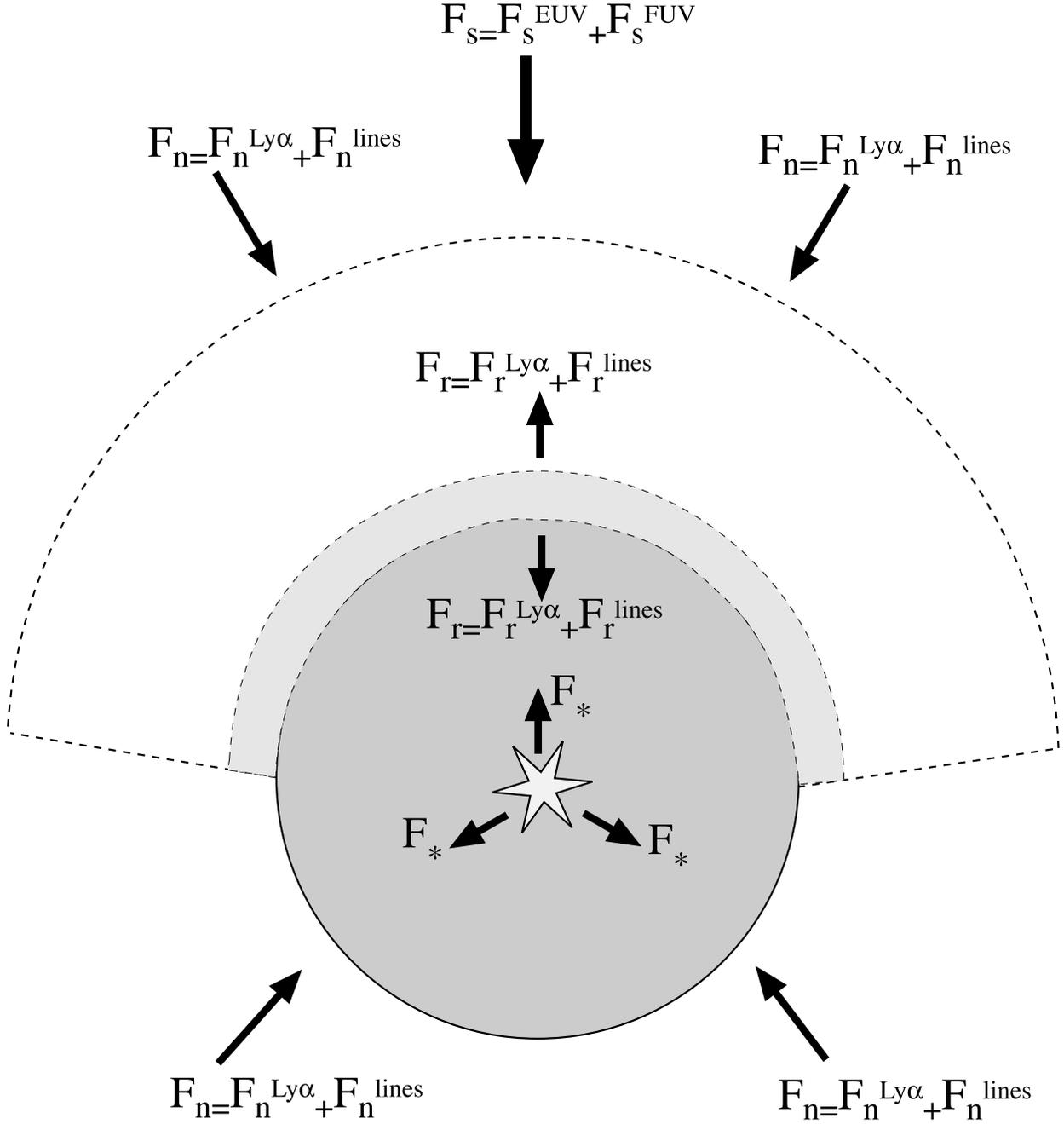}
%\plottwo{EPSFILE}{EPSFILE}
%\plotfiddle{EPSFILE}{VSIZE}{ROT}{HSF}{VSF}{HTRANS}{VTRANS}
\caption{Sketch of our assumed globule geometry with the radiative fluxes relevant for 
the dust heating.
\label{fluxes_in_the_globule}}
\end{figure}

\clearpage
\newpage
\begin{figure} %3
%\figurenum{TEXT}
%\epsscale{NUM}
\plotone{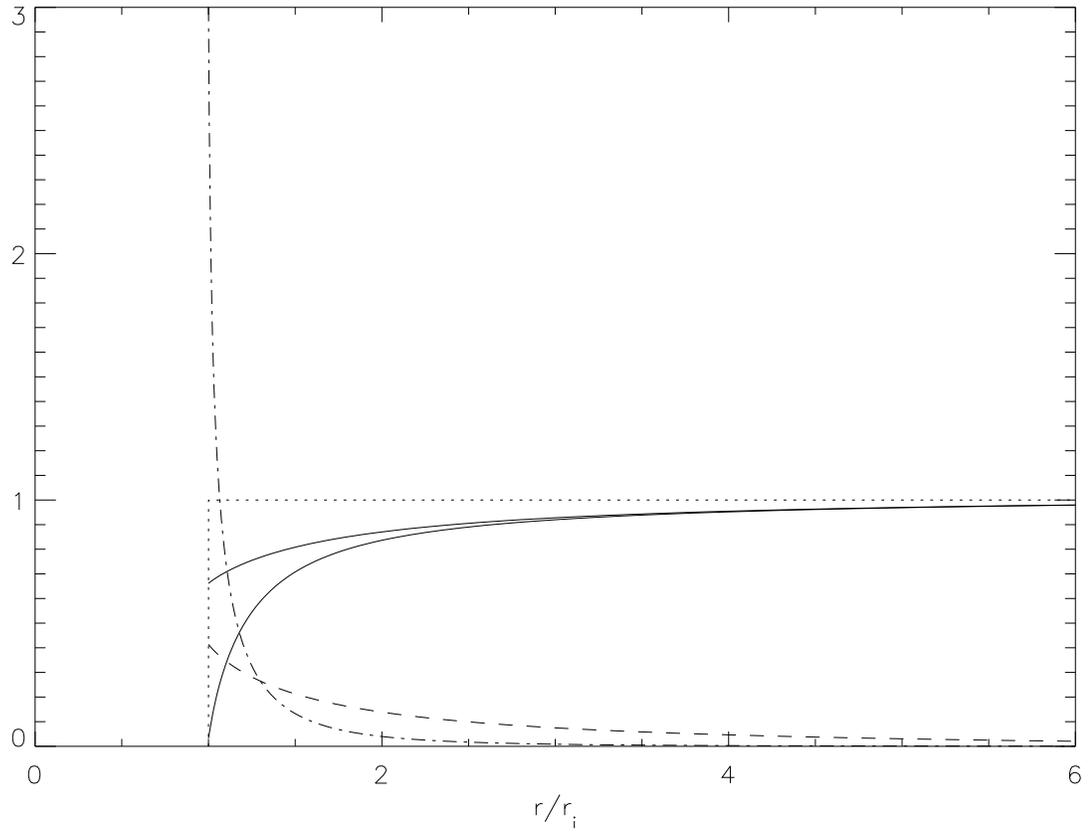}
%\plottwo{EPSFILE}{EPSFILE}
%\plotfiddle{EPSFILE}{VSIZE}{ROT}{HSF}{VSF}{HTRANS}{VTRANS}
\caption{Propagation of the flux from \thC\, within the ionized wind. The two
solid lines represent the FUV flux (upper line) and the EUV flux (lower line).
The dashed line represents the dust optical depth, whereas the dot-dashed line 
represents the gas optical depth. The dotted line represents the ionization fraction.
 \label{thinshell}}
\end{figure}

\newpage
\begin{figure} %4
%\figurenum{TEXT}
%\epsscale{NUM}
\plotone{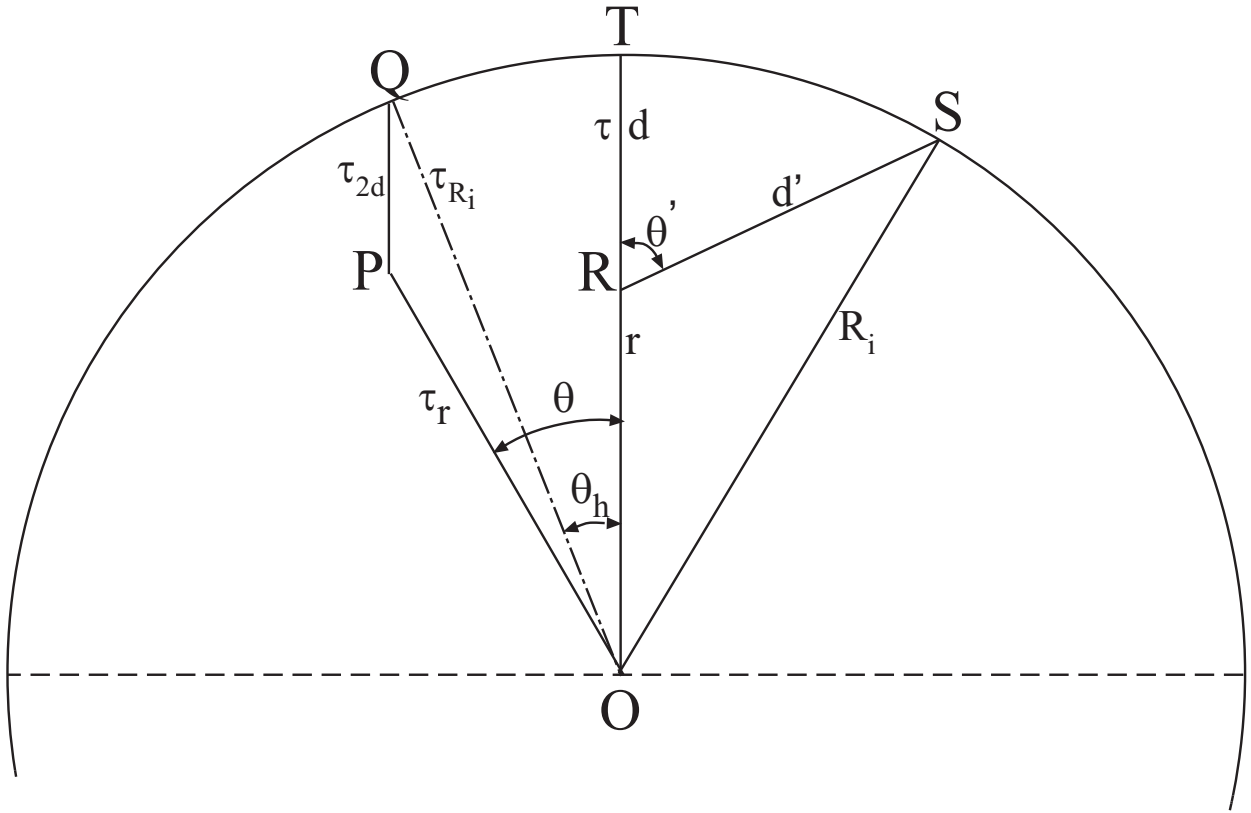}
%\plottwo{EPSFILE}{EPSFILE}
%\plotfiddle{EPSFILE}{VSIZE}{ROT}{HSF}{VSF}{HTRANS}{VTRANS}
\caption{Globule geometry. The left side shows our naming conventions for the optical depths, the right side
for the distances. There is a simple relation between optical depth and distance (Equation~18), and
corresponding quantities share the same index. Note, however, that $\tau$ corresponds to the distance $R_i-r$. \label{globule_geometry}}
\end{figure}

\newpage
\begin{figure} %5
%\figurenum{TEXT}
%\epsscale{NUM}
\plotone{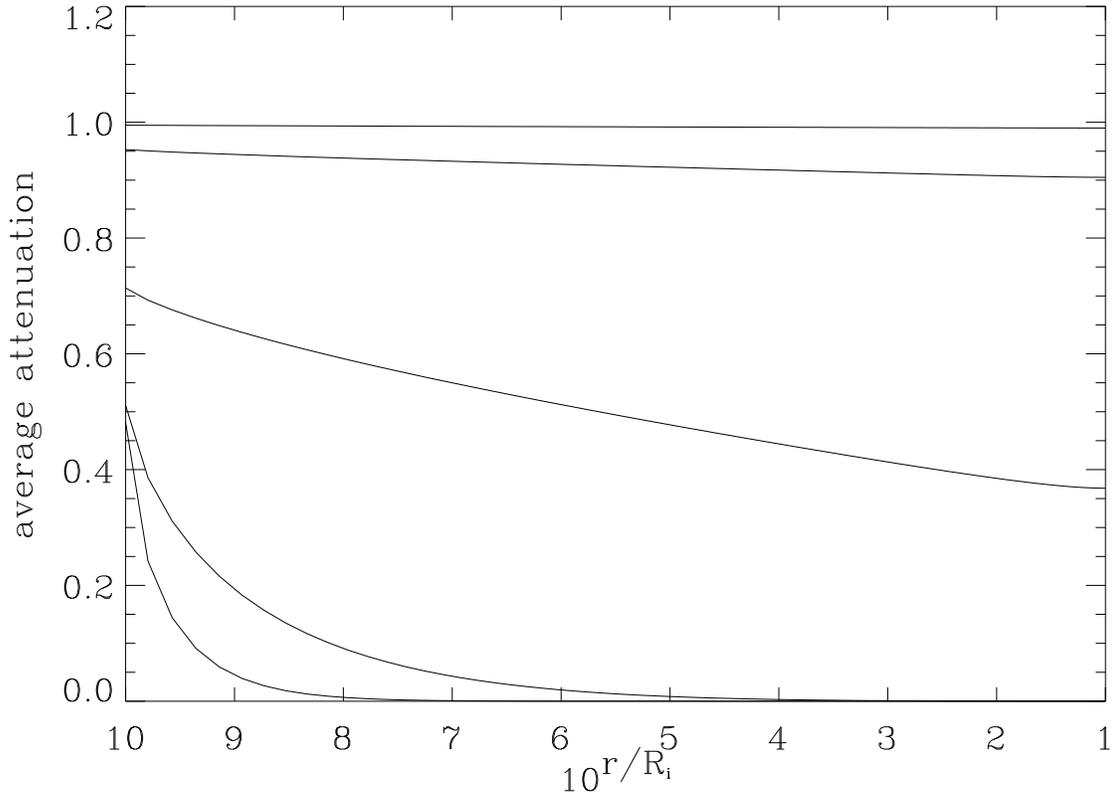}
%\plottwo{EPSFILE}{EPSFILE}
%\plotfiddle{EPSFILE}{VSIZE}{ROT}{HSF}{VSF}{HTRANS}{VTRANS}
\caption{Attenuation of an isotropic radiation field within a homogeneous spherical globule with radial optical depth $\tau_R$. Fom top to the bottom: $\tau_{R_i}$=0.01, 0.1, 1, 10, 30.
\label{Lyalphainglobule_NEW_2}}
\end{figure}

\newpage
\begin{figure} %6
%\figurenum{TEXT}
%\epsscale{NUM}
\plotone{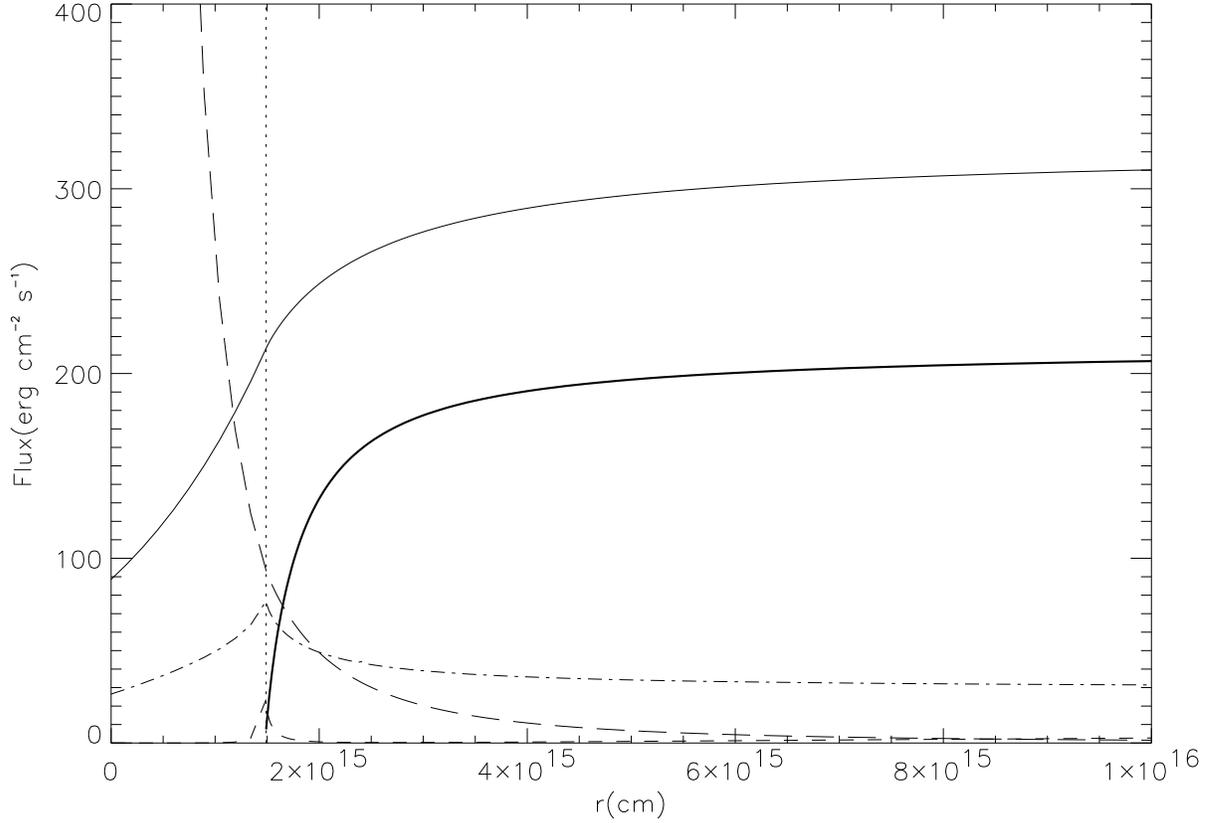}
%\plottwo{EPSFILE}{EPSFILE}
%\plotfiddle{EPSFILE}{VSIZE}{ROT}{HSF}{VSF}{HTRANS}{VTRANS}
\caption{Flux density along the globule axis. The vertical dotted line indicates the position of the
ionization front at $R_i = 100$~AU. The fluxes represented are: non-ionizing FUV
radiation from \thC\, (solid thin line), ionizing EUV  radiation from \thC\, (solid thick line); radiation from the globule central star (long-dashed line), recombination line radiation from the nebular environment and from the ionization front (dot-dashed line); \Lya\, radiation from the nebular environment and from the ionization front(short-dashed line), this last close to zero
and mostly visible with a peak at the ionization front.
\label{Fshell_theta1C}}
\end{figure}

\newpage
\begin{figure} %7
%\figurenum{TEXT}
%\epsscale{NUM}
\plotone{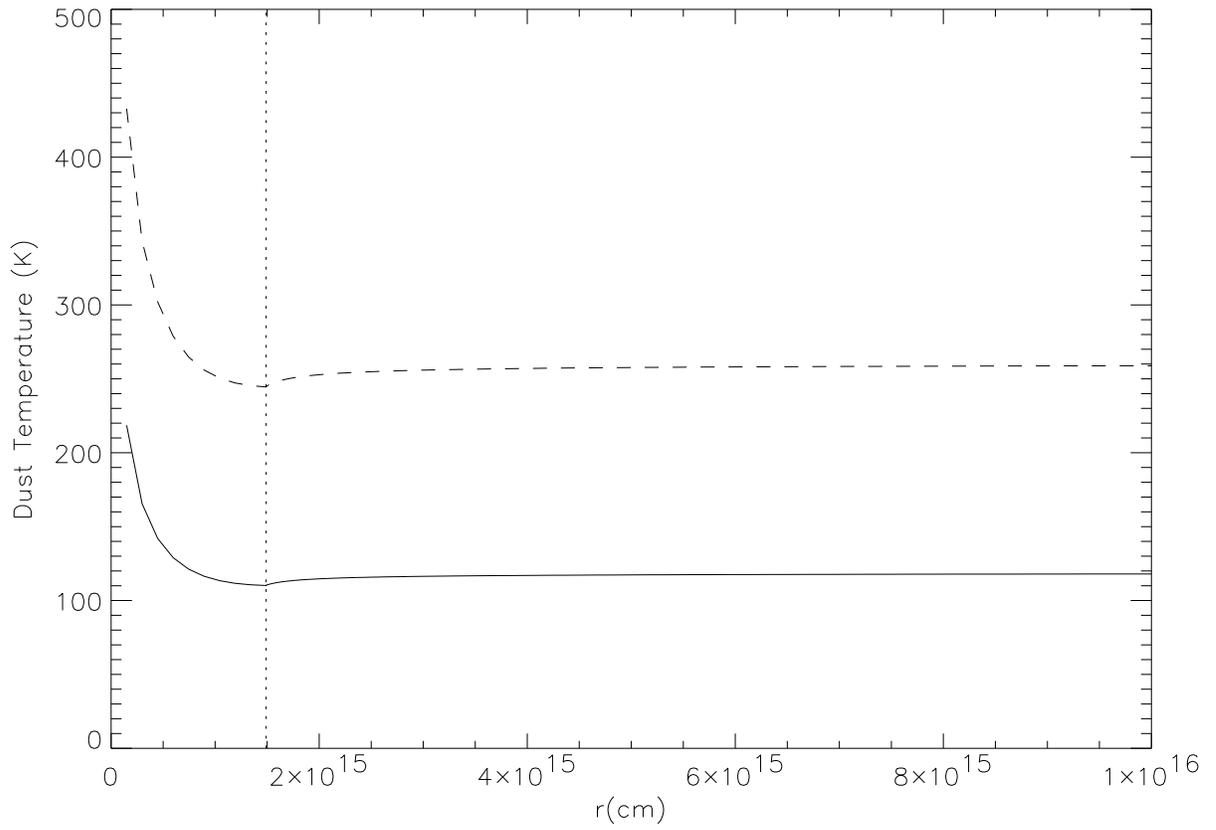}
%\plottwo{EPSFILE}{EPSFILE}
%\plotfiddle{EPSFILE}{VSIZE}{ROT}{HSF}{VSF}{HTRANS}{VTRANS}
\caption{Temperature distribution along the globule axis resulting from the
radiation sources represented in Figure~\ref{Fshell_theta1C}.
\label{Tglobule_axis}}
\end{figure}

\newpage
\begin{figure} %8
%\figurenum{TEXT}
%\epsscale{NUM}
\plotone{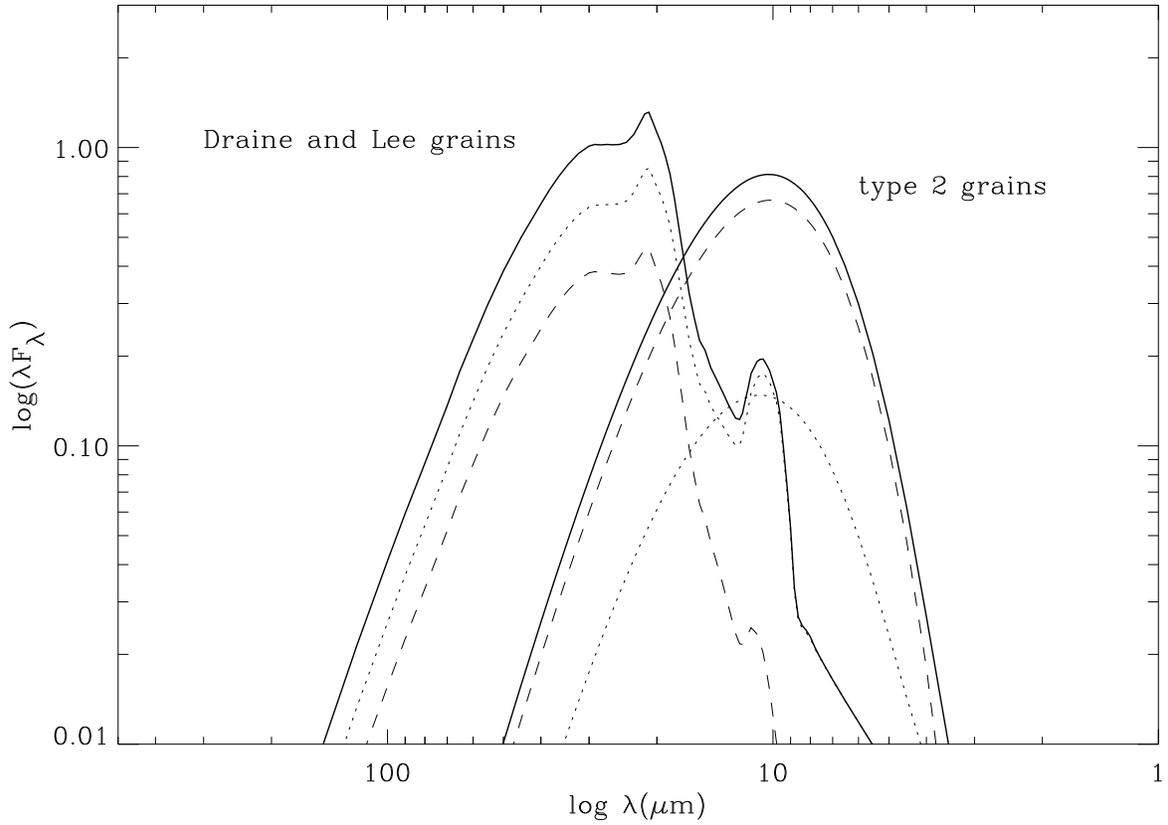} 
%\plottwo{EPSFILE}{EPSFILE}
%\plotfiddle{EPSFILE}{VSIZE}{ROT}{HSF}{VSF}{HTRANS}{VTRANS}
\caption{Spectrum emitted by a dusty globule with our standard set of parameters. The two solid curves are relative to type-1 grains (left) and type-2 grains (right). Each curve is the sum of   the contributions from the neutral (dashed line) and ionized (dotted line) parts of the envelope. Self absorpion has not been considered. The stellar spectrum is not plotted.
\label{IR_FLUX_envelopeonly}}
\end{figure}

\newpage
\begin{figure} % 9
%\figurenum{TEXT}
%\epsscale{NUM}
\plotone{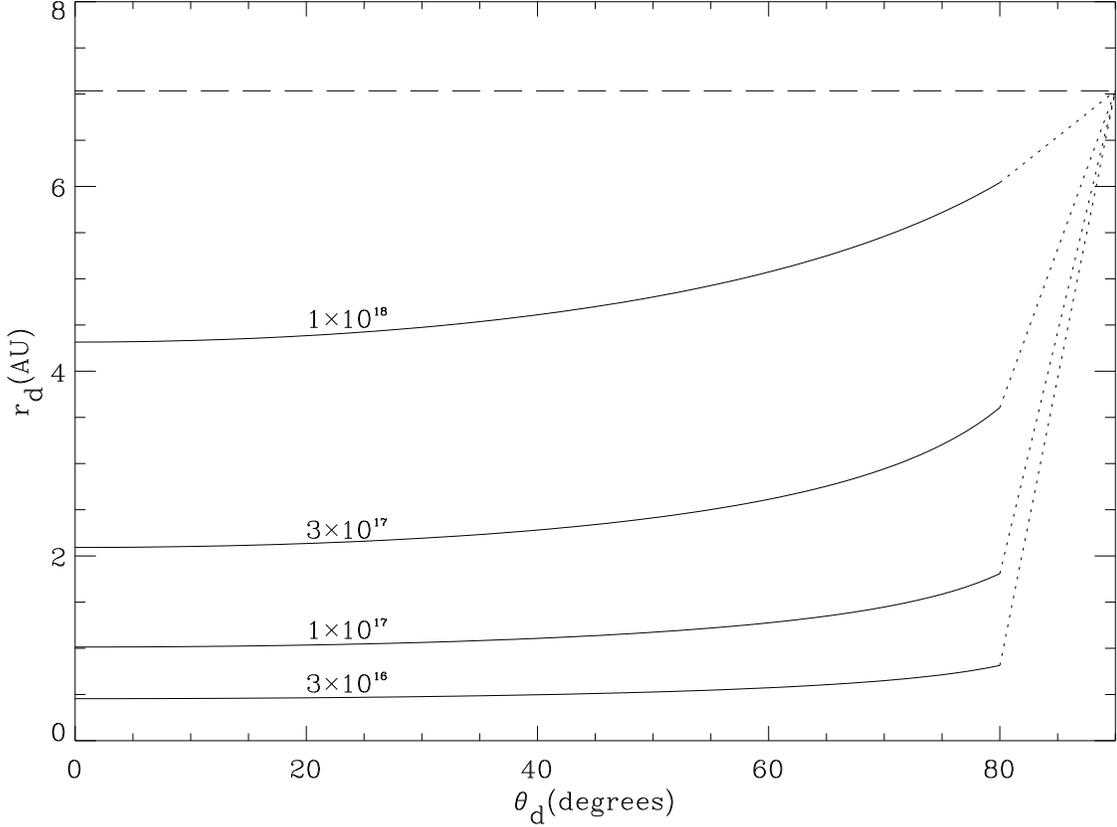}
%\plottwo{EPSFILE}{EPSFILE}
%\plotfiddle{EPSFILE}{VSIZE}{ROT}{HSF}{VSF}{HTRANS}{VTRANS}
\caption{Radius $a_d$ at which the disk heating from the central star equals that from the HII region as a 
function of the disk tilt angle $\theta_d$ and of the distance from \thC. 
From top to bottom, the lines represent decreasing distances from \thC. The dashed part 
of each line represents the case of pure nebular heating, occurring on the
rear face of the disk (dark silhouette) within the HII region. The
dotted lines at $\theta_d=80^\circ-90^\circ$ are relative to the case of a disk aligned 
almost edge-on with respect to
the direction of \thC. The stellar parameters are $T_\star=4,500$~K and $L_\star=2.5 L_\odot$.
\label{Inoutdiskface}}
\end{figure}

\newpage
\begin{figure} % 10
%\figurenum{TEXT}
%\epsscale{NUM}
\plotone{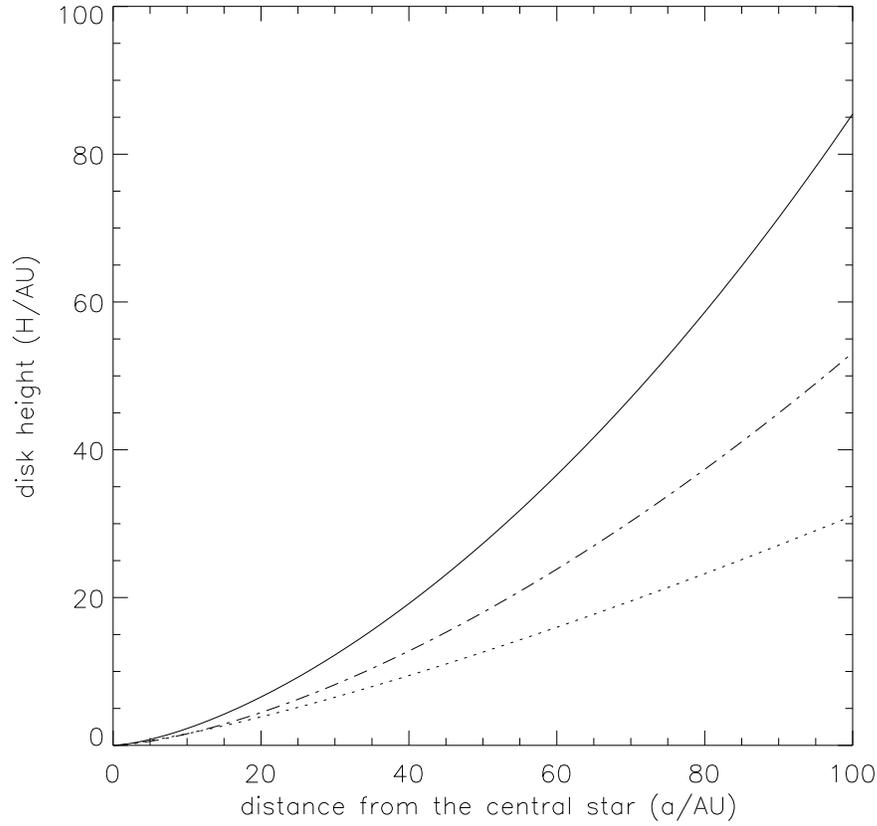}
%\plottwo{EPSFILE}{EPSFILE}
%\plotfiddle{EPSFILE}{VSIZE}{ROT}{HSF}{VSF}{HTRANS}{VTRANS}
\caption{Vertical sections of the disk. The continuous line 
represents the locus of the disk photosphere when both the flux from \thC\, and the
nebular radiation are taken into account. The dot-dashed line has been obtained by 
neglecting the flux from \thC\, and is therefore representative of the dark face
of the disk. The dotted line has been obtained taking into account only the 
radiation from the disk central star (as in CG97).
\label{alpha}}
\end{figure}

\newpage
\begin{figure} % 11
%\figurenum{TEXT}
%\epsscale{NUM}
\plotone{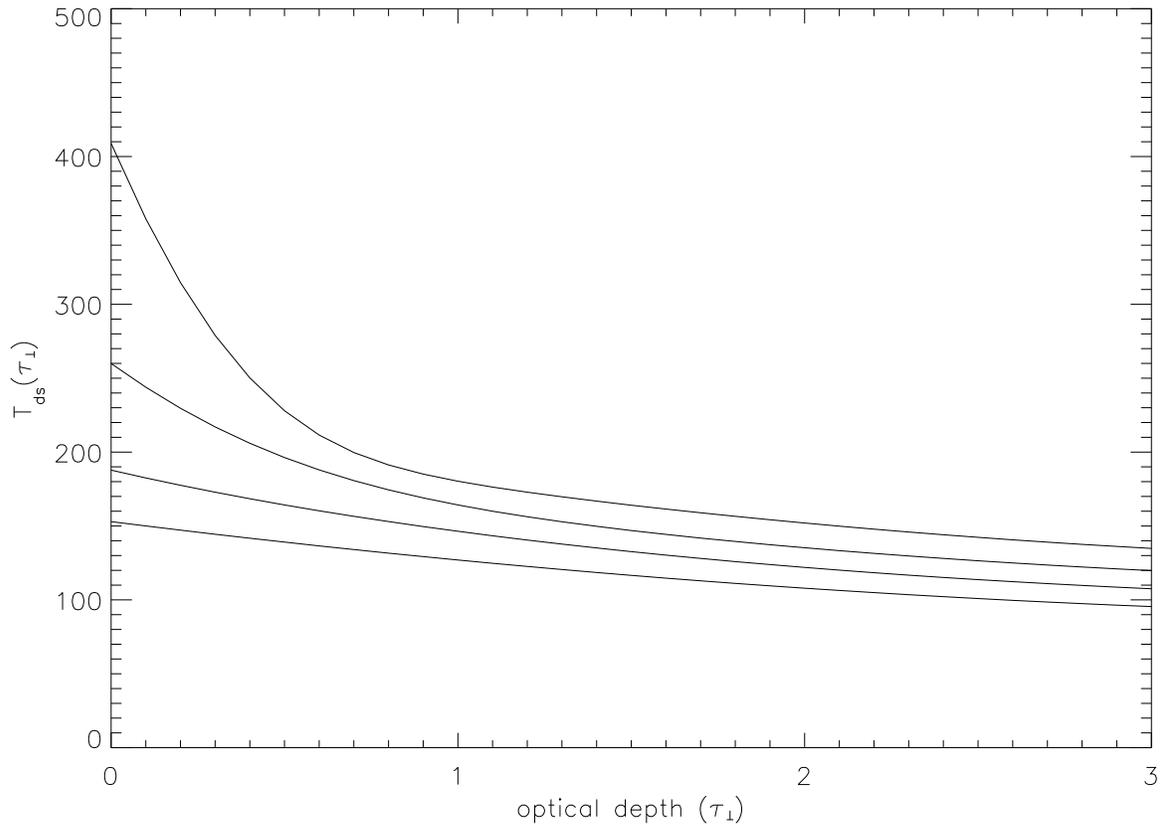}
%\plottwo{EPSFILE}{EPSFILE}
%\plotfiddle{EPSFILE}{VSIZE}{ROT}{HSF}{VSF}{HTRANS}{VTRANS}
\caption{Temperature of the superheated layer in absence of external luminosity at distances (top to bottom) a=3, 10, 30, 100 AU from the central star. \label{Tds_NOENVELOPE}}
\end{figure}

\newpage
\begin{figure} %12
%\figurenum{TEXT}
%\epsscale{NUM}
\plotone{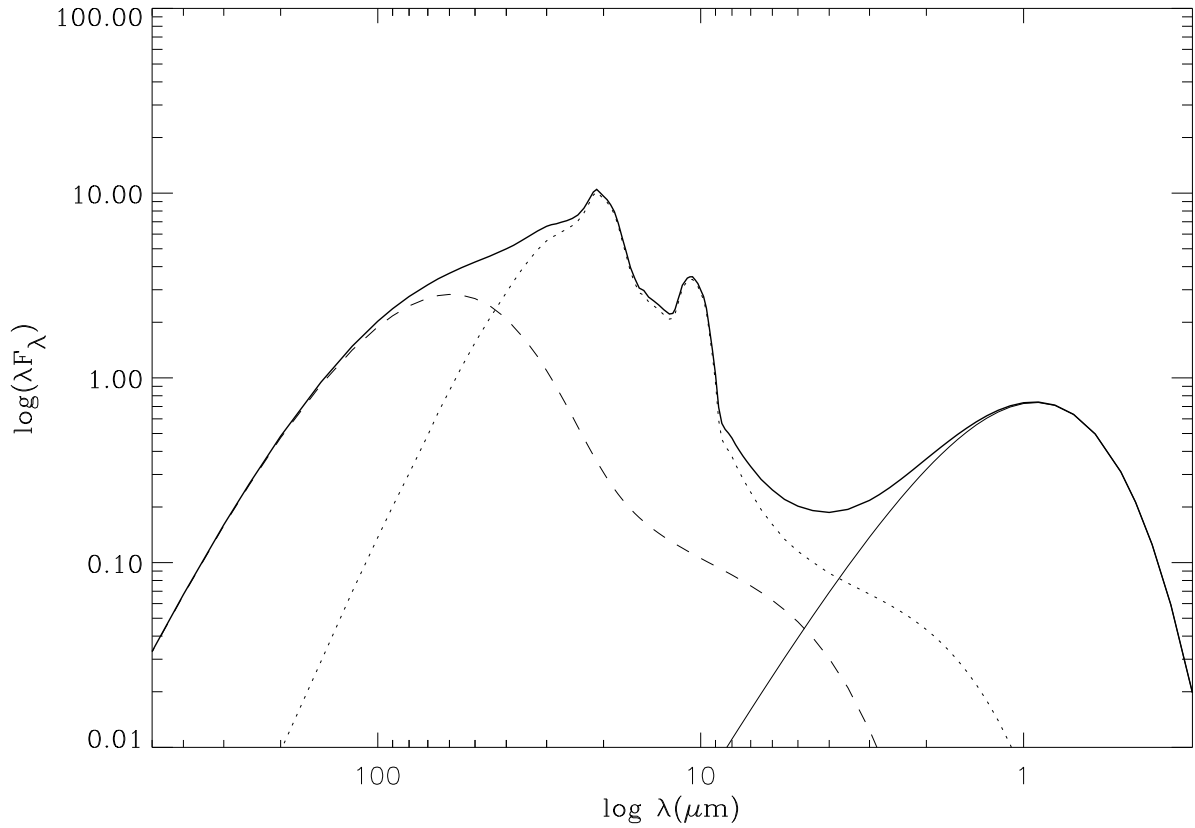}
%\plotfiddle{EPSFILE}{VSIZE}{ROT}{HSF}{VSF}{HTRANS}{VTRANS}
\caption{SED for a hydrostatic, radiative equilibrium disk directly exposed to the flux from \thC. Solid line: stellar photosphere; dotted line:
superheated layer; dashed line: inner disk. The thick solid line represents the total SED. The standard set of parameters with Draine and Lee (1984) silicates has been assumed.
\label{Ldisk_NOENVELOPE_front}}
\end{figure}

\newpage
\begin{figure} %13
%\figurenum{TEXT}
%\epsscale{NUM}
%\plotone{EPSFILE}
\plotone{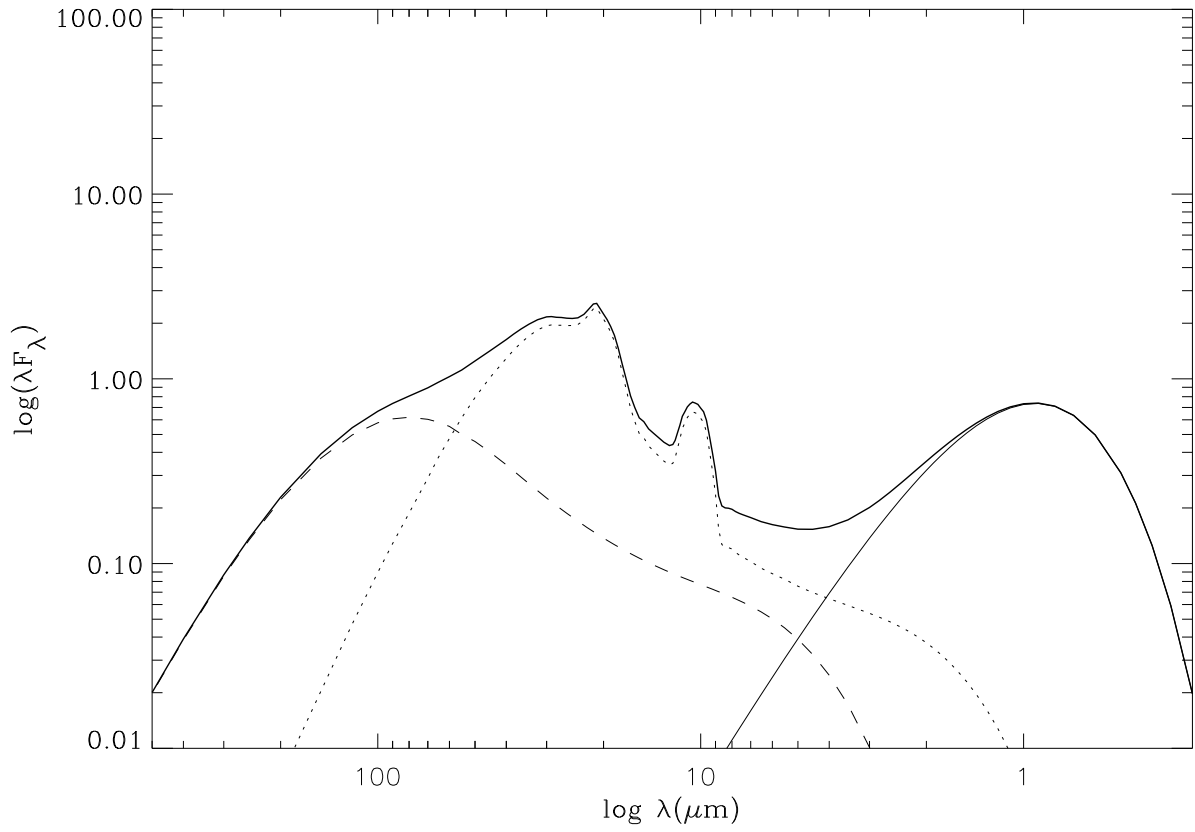}
%\plotfiddle{EPSFILE}{VSIZE}{ROT}{HSF}{VSF}{HTRANS}{VTRANS}
\caption{Same as Figure~\ref{Ldisk_NOENVELOPE_front}, for a hydrostatic, radiative equilibrium disk directly exposed to the nebular radiation only. 
\label{Ldisk_NOENVELOPE_back}}
\end{figure}

\clearpage

\newpage
\begin{figure} %14
%\figurenum{TEXT}
%\epsscale{NUM}
\plotone{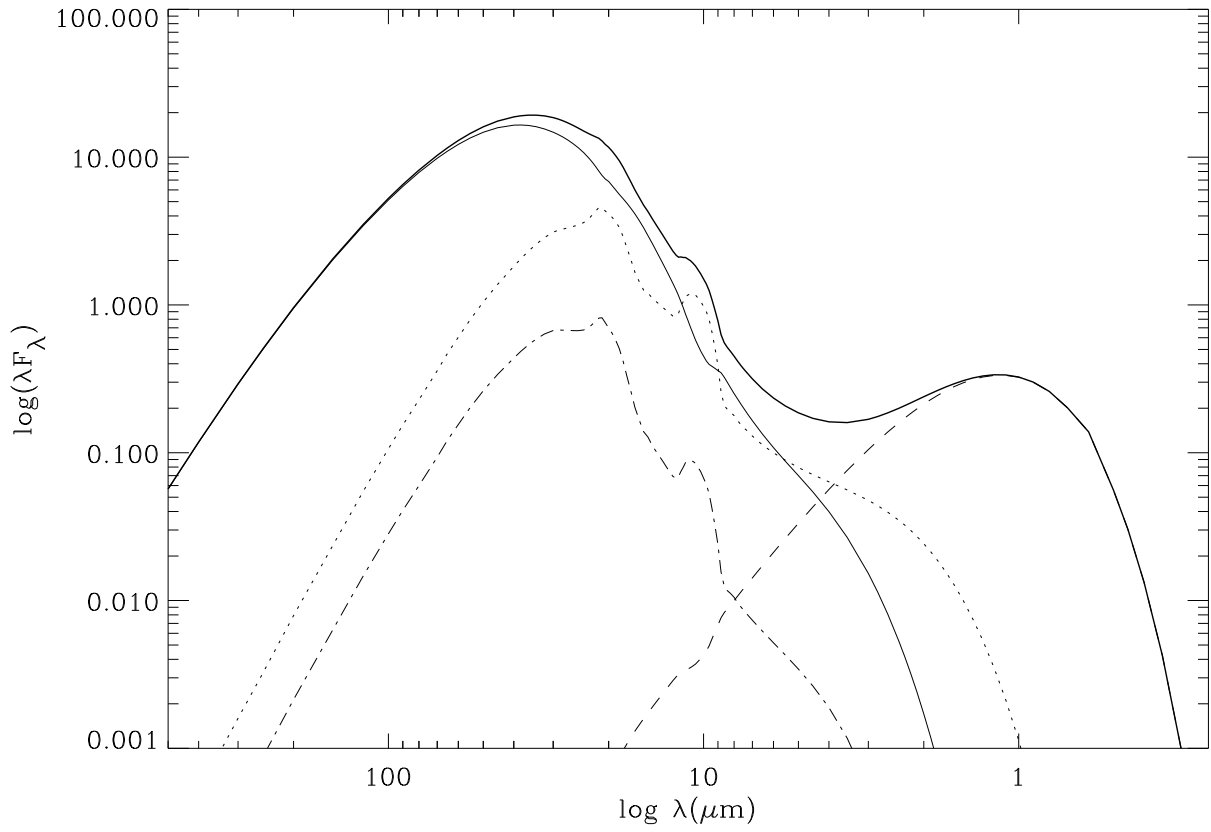}
%\plottwo{EPSFILE}{EPSFILE}
%\plotfiddle{EPSFILE}{VSIZE}{ROT}{HSF}{VSF}{HTRANS}{VTRANS}
\caption{SED of a flaring disk embedded within a photoevaporated envelope. Solid line: stellar photosphere; dotted line: superheated layer; 
dashed line: inner disk; dot-dashed line: outer envelope; thick solid line: total SED. 
\label{Ldisk+envelope_ALL}}
\end{figure}

\newpage
\begin{figure} %15
%\plotone{./f15.eps}
%\plotone{./SC3b.eps}
\plotone{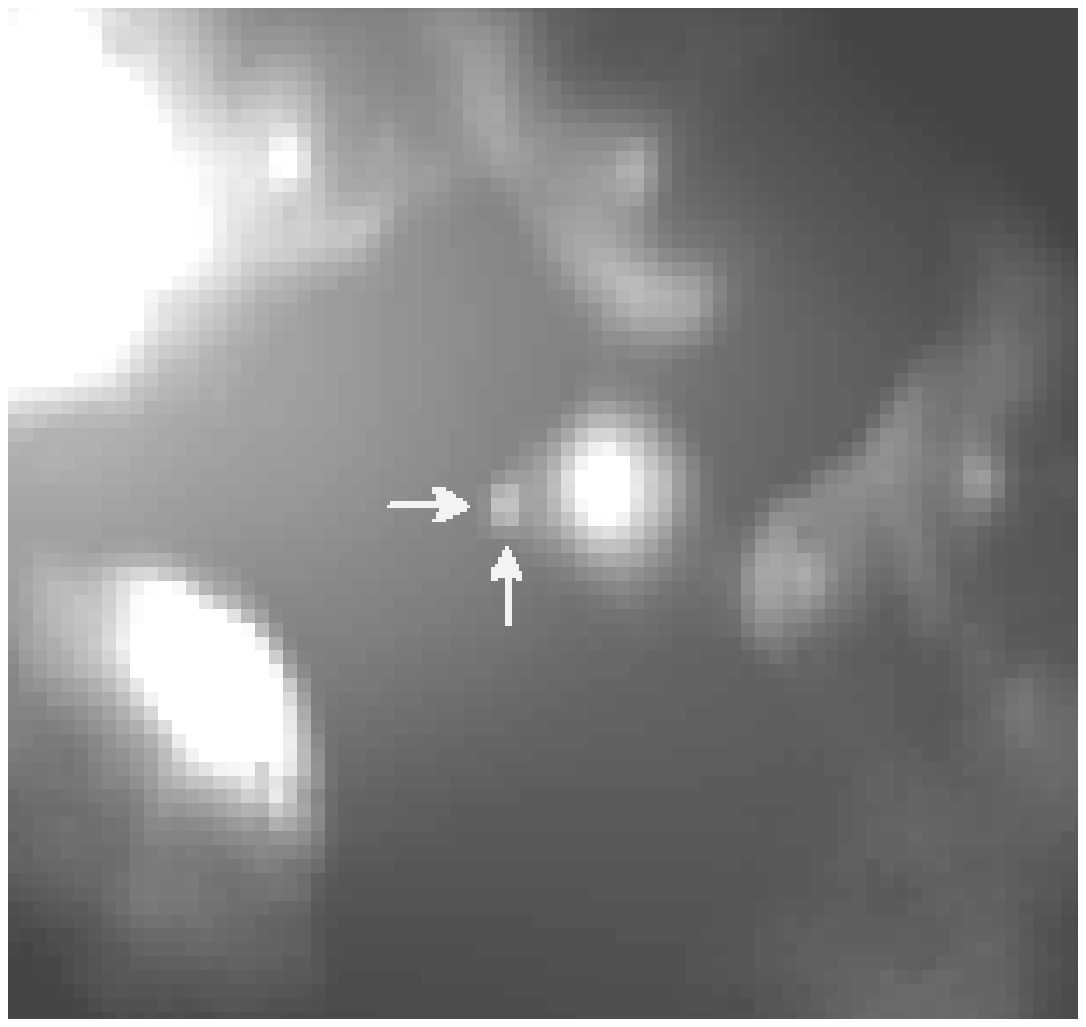}
\caption{Image at 10\,\micron\, of the \thC\, field. The arrows mark the position of \thC. Source SC3 is the bright knot at located $\approx 1\arcsec$ to the West of \thC. The field size is approximately $15\arcsec\times15\arcsec$ and the pixel size is $0.26\arcsec$/pixel. The image has been obtained with the MAX camera on the UKIRT telescope and is part of a large mosaic of the Orion nebula core (Robberto et al 2002, in preparation).
\label{th1C}}
\end{figure}

\newpage
\begin{figure} %16
%\figurenum{TEXT}
%\epsscale{NUM}
\plotone{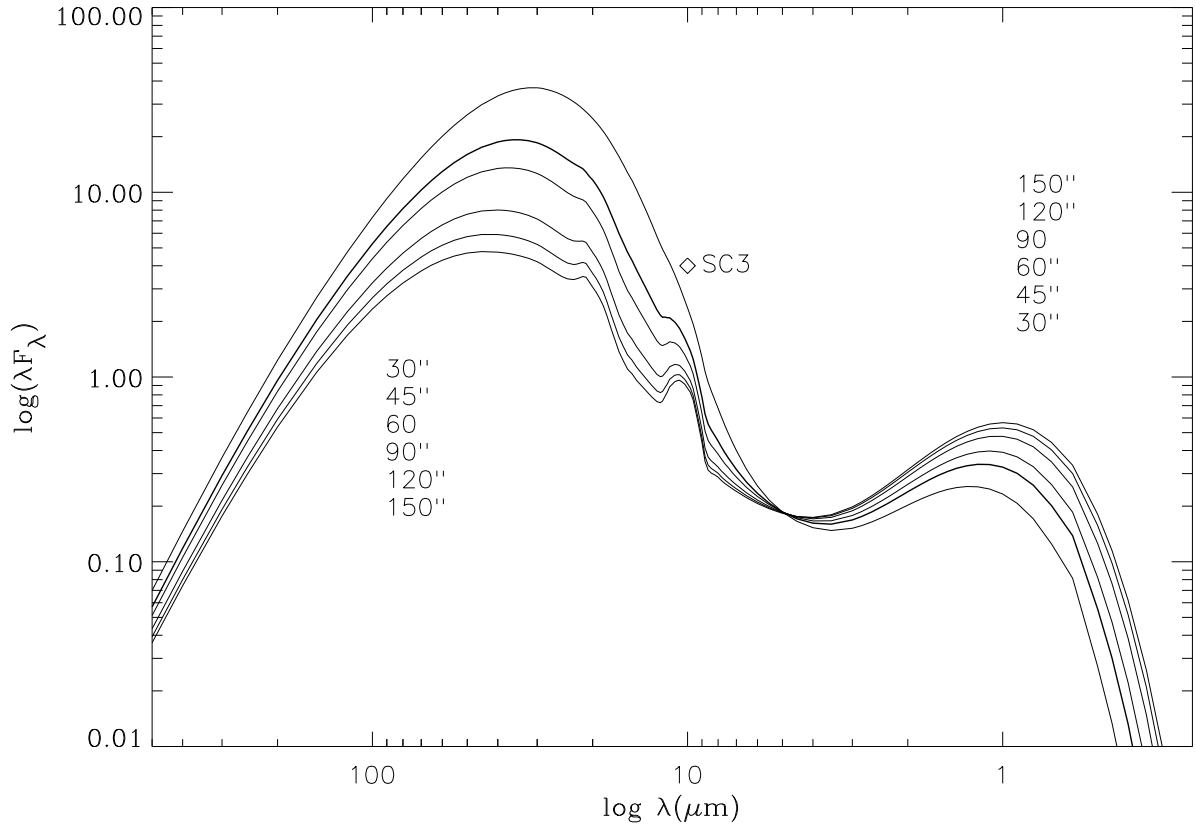}
%\plottwo{EPSFILE}{EPSFILE}
%\plotfiddle{EPSFILE}{VSIZE}{ROT}{HSF}{VSF}{HTRANS}{VTRANS}
\caption{SED of a flaring disk embedded within a photoevaporated envelope for different values of the tilt angle with respect to the Earth $\theta_\earth$. The thick line refers to the standard case. 
\label{tiltangle}}
\end{figure}

%\newpage
\begin{figure} %17
%\figurenum{TEXT}
%\epsscale{NUM}
\plotone{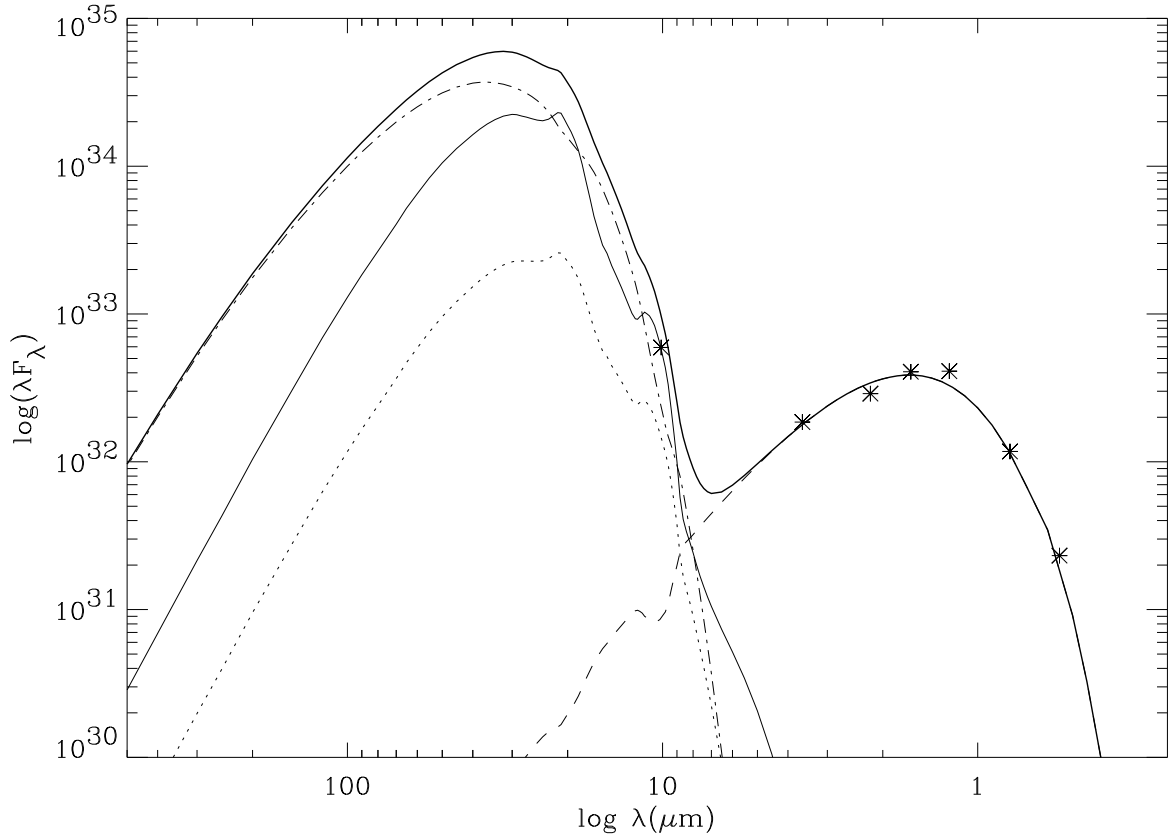}
%\plottwo{EPSFILE}{EPSFILE}
%\plotfiddle{EPSFILE}{VSIZE}{ROT}{HSF}{VSF}{HTRANS}{VTRANS}
\caption{Photometry of the proplyd source 177-341 = HST~1 (crosses) and model
SED assuming a flaring disk embedded within a photoevaporated envelope.
See text for the fit parameters. The 10\,\micron\, photometric point derives from the image presented in Figure~\ref{4proplyds}. PHotometry at shorter wavelengths is from Hillnebrand (1997) and from Robberto et al. (2002).
\label{datacomparison}}
\end{figure}

%\newpage
\begin{figure} %18
%\figurenum{TEXT}
%\epsscale{NUM}
\plotone{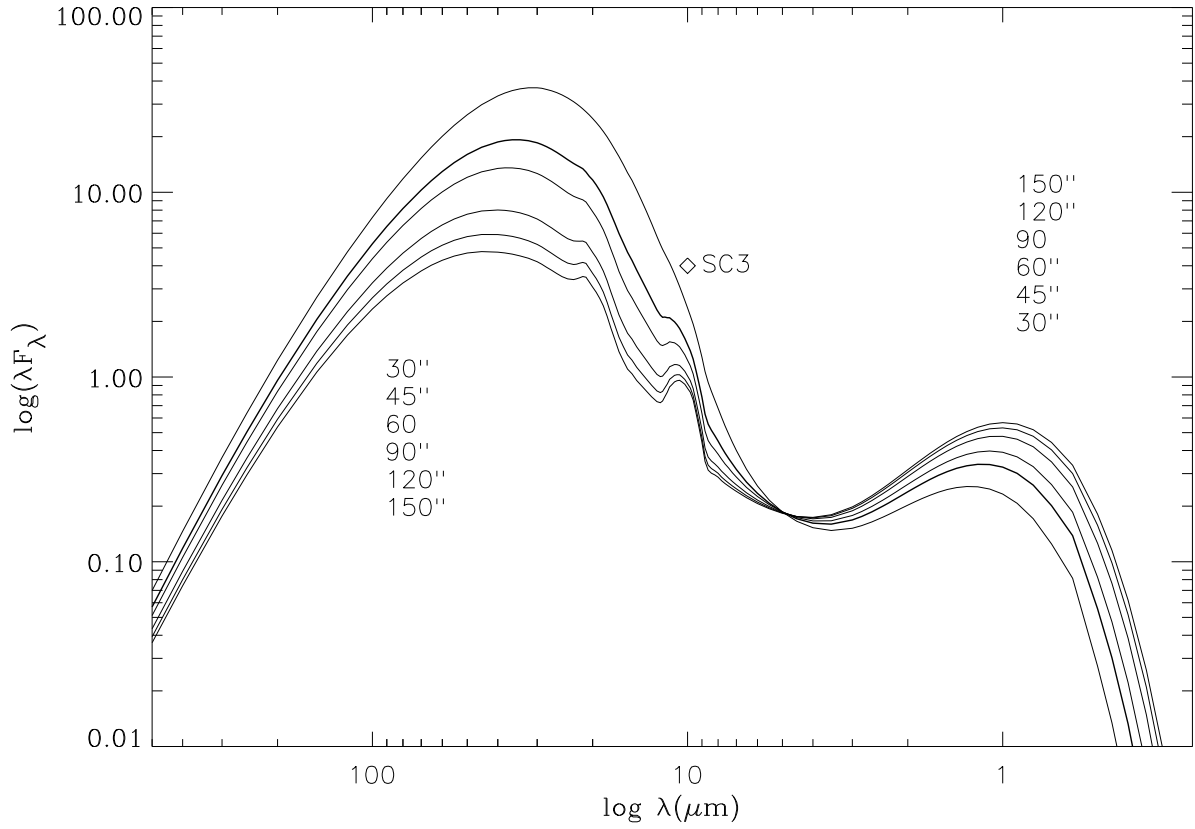}
%\plottwo{EPSFILE}{EPSFILE}
%\plotfiddle{EPSFILE}{VSIZE}{ROT}{HSF}{VSF}{HTRANS}{VTRANS}
\caption{SED of a flaring disk embedded within a photoevaporated envelope for different values of the distance from \thC. The thick line refers to the standard case.
\label{SEDcompare_distance}}
\end{figure}

%\newpage
\begin{figure} %19
%\figurenum{TEXT}
%\epsscale{NUM}
\plotone{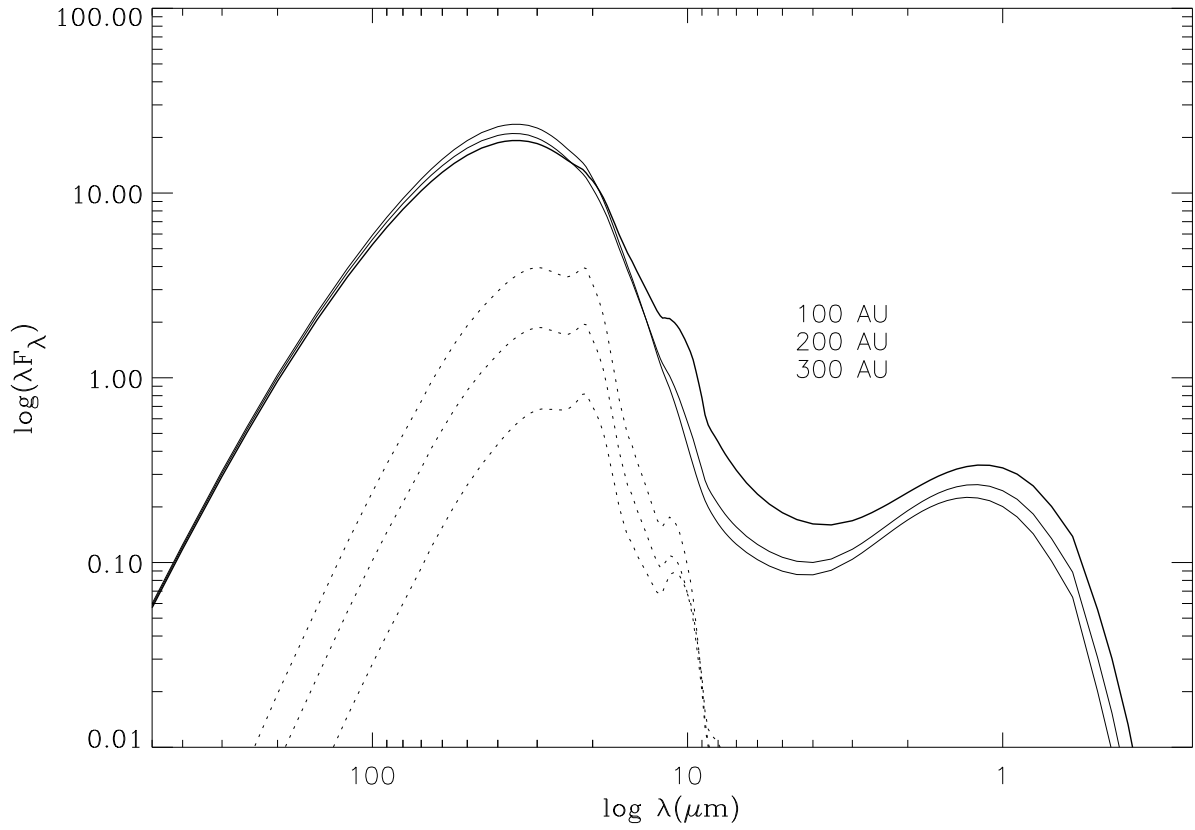}
%\plottwo{EPSFILE}{EPSFILE}
%\plotfiddle{EPSFILE}{VSIZE}{ROT}{HSF}{VSF}{HTRANS}{VTRANS}
\caption{SED of a flaring disk embedded within a photoevaporated envelope for different values of the ionization radius $R_{out}$.
\label{SEDcompare_Rout}}
\end{figure}

%\newpage
\begin{figure} %20
%\figurenum{TEXT}
%\epsscale{NUM}
\plotone{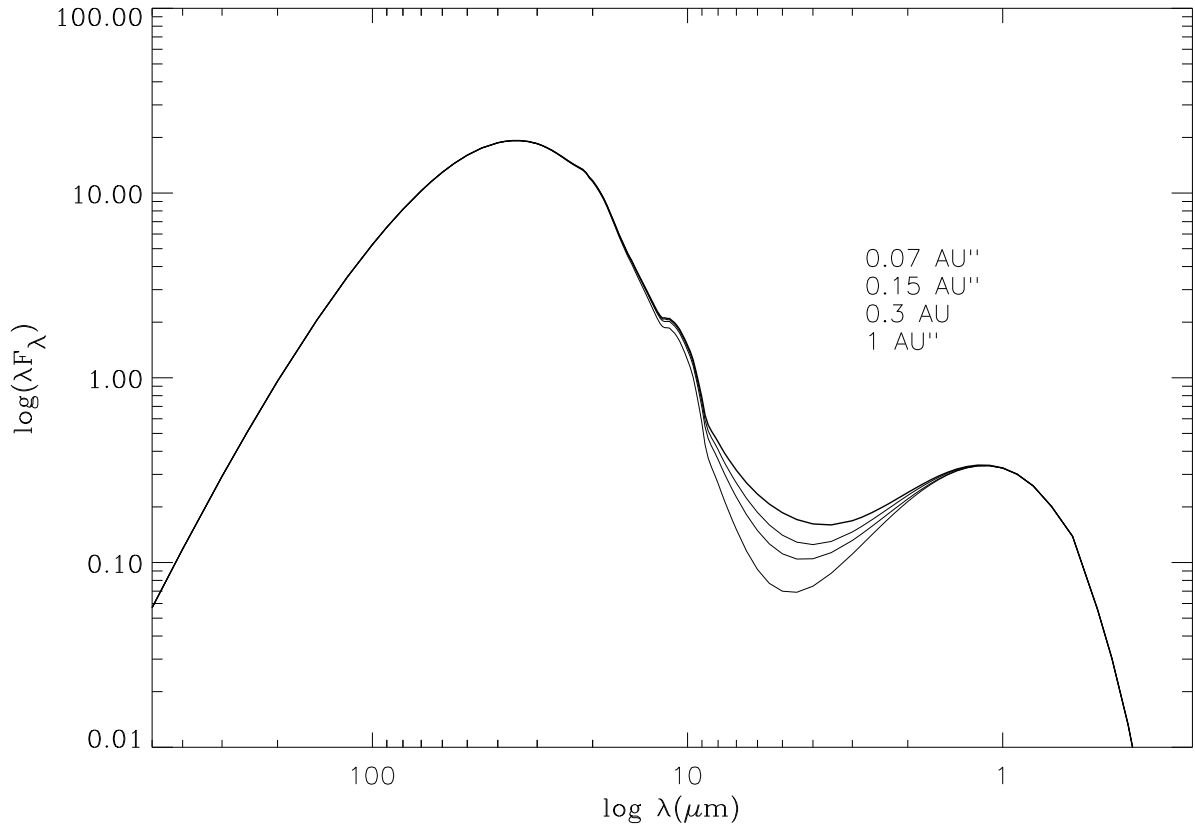}
%\plottwo{EPSFILE}{EPSFILE}
%\plotfiddle{EPSFILE}{VSIZE}{ROT}{HSF}{VSF}{HTRANS}{VTRANS}
\caption{SED of a flaring disk embedded within a photoevaporated envelope for
different values of the disk inner radius $a_i$. 
\label{SEDcompare_ai}}
\end{figure}

%\newpage
\begin{figure} %21
%\figurenum{TEXT}
%\epsscale{NUM}
\plotone{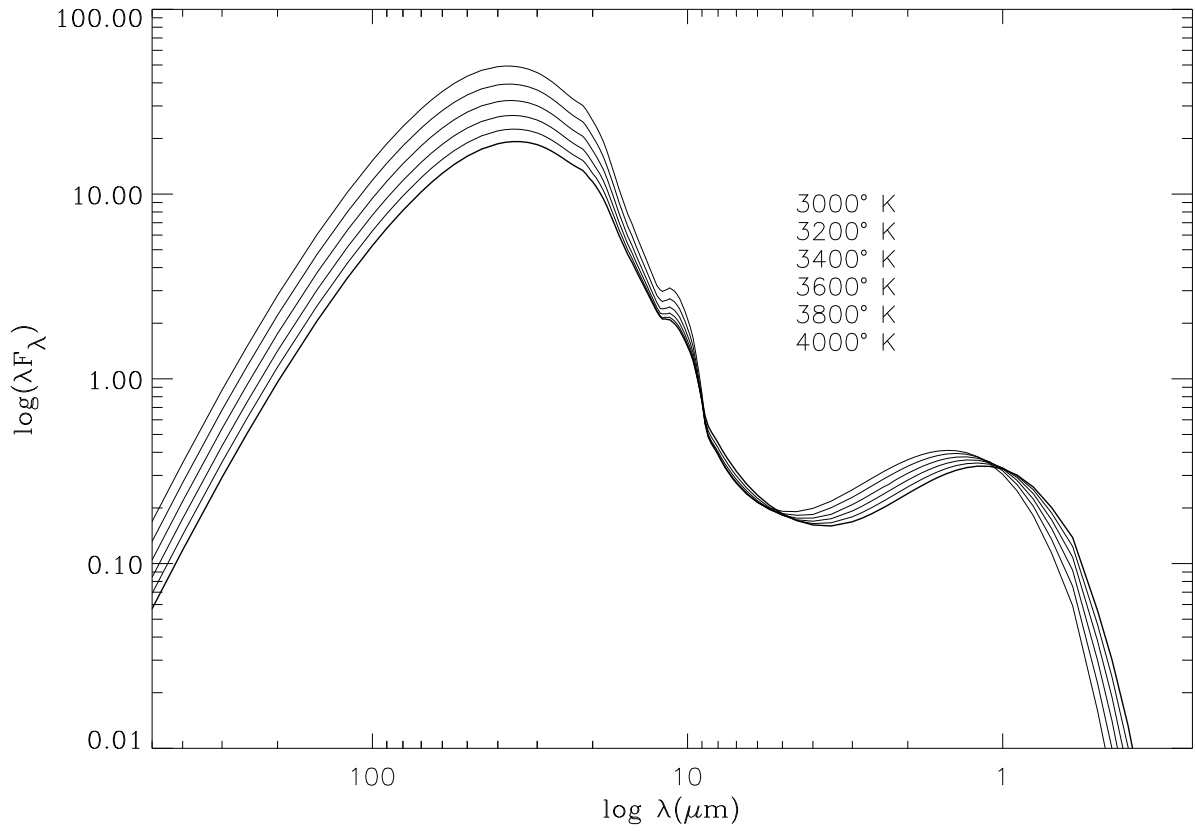}
%\plottwo{EPSFILE}{EPSFILE}
%\plotfiddle{EPSFILE}{VSIZE}{ROT}{HSF}{VSF}{HTRANS}{VTRANS}
\caption{SED of a flaring disk embedded within a photoevaporated envelope for different values of the 
effective stellar temperature $T_\star$.
\label{SEDcompare_Tstar}}
\end{figure}

%\newpage
\begin{figure} %22
%\figurenum{TEXT}
\epsscale{0.80}
\plotone{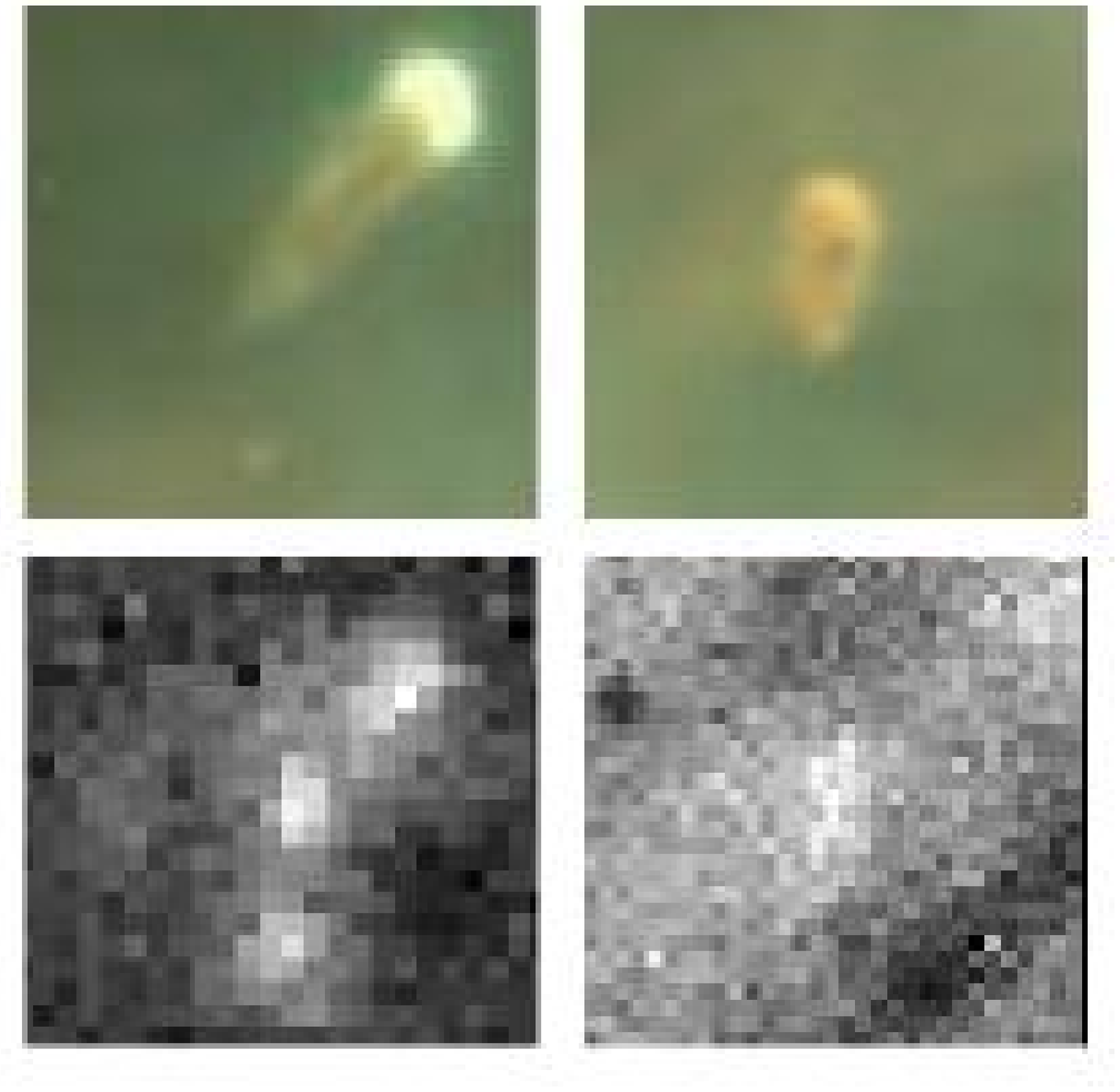}
%\plottwo{EPSFILE}{EPSFILE}
%\plotfiddle{EPSFILE}{VSIZE}{ROT}{HSF}{VSF}{HTRANS}{VTRANS}
\caption{The prominent proplyd sources 177-341 (HST~1, left) and 182-413 
(HST~10, right) imaged by the HST (top) and MAX on UKIRT (bottom). 
The HST images refer to a combination of recombination lines, 
namely [NII] (red), H$\alpha$ (green) and [OIII] (blue). 
The MAX images have been taken in N-broad band with integration time 
of the order of 600 seconds on source (see Robberto et al. 2002 for details)
\label{4proplyds}}
\end{figure}


\begin{references}
\reference{Bally+98a}
	Bally, J., Sutherland, R. S., Devine, D., Johnstone, D. 1998, \aj, 116, 293 (1998a) 
\reference{Bally+98b}
	Bally, J., Testi, L., Sargent, A., \& Carlstrom, J. 1998, \aj, 116, 854 (1998b)
\reference{Bally+00}
	Bally, J., O'Dell, C. R., \& McCaughrean, M. J. 2000, \aj, 119, 2919
\reference{Beckwith+90}
	Beckwith, S. V. W., Sargent, A. I., Chini, R. S., \& Gusten, R. 1990, \aj, 99, 924
\reference{SVWBAS96}
	Beckwith, S. V. W., Sargent, A. I.  1996, Nature, 383, 139
\reference{Butner+94}
	Butner, H. M., Natta, A., and Evans, N. J. II 1994, \apj, 420, 326 
%\reference{Calvet+92}
%Calvet, N., Magris, G., Patino, A., D'Alessio, P. 1992, Rev. Mex. Atron. Astrof., 24, 27
\reference{CG97}
	Chiang, E. I., Goldreich, P. 1997, \apj, 490, 368 
\reference{CG99}
	Chiang, E. I., Goldreich, P. 1999, \apj, 519, 279 
\reference{CH99}
	Choi, P. \& Herbst, W. 1996, \aj, 111, 283 
\reference{Draine+Lee84}
	Draine, B. T., \& Lee, H. M. 1984, ApJ, 285, 89
\reference{Dyson68}
	Dyson, J. E. 1968, \apjs, 1, 388
\reference{Edwards+93}
	Edwards, S., Strom, S., Hartigan, P., Strom, K. M., Hillenbrand, L. A.,  Herbst, W., Attridge, J., Merril, K. M., Probst, R., \& Gatley, I. 1993, \aj, 106,  372
%\reference{Harwit}
%	 Harwit, Astrophysical Concepts, Eq. 4-23
\reference{Hartmann+98}
	Hartmann, K. 1998, Accretion Processes in Astrophysics, Cambridge University Press, 157
\reference{Hayward+94}
	Hayward, T. L., Houck, J. R., Miles, J. W. 1994, \apj, 433, 157
\reference{Hayward+MJM}
	Hayward, T. L., McCaughrean, M. J. 1997, \aj, 113, 346
\reference{Henney+96}
	Henney, W. J., Raga, A. C., Lizano, S., \& Curiel, S. 1996, \apj, 465, 216
\reference{Hillenbrand97}
	Hillenbrand, L. A. 1997,  \aj, 113, 1733
\reference{Hummer+Kunasz80}
	Hummer, D. G., \& Kunasz, P. B. 1980, \apj, 236, 609
\reference{Johnstone+98}
	Johnstone, D., Hollenbach, D., \& Bally, J. 1998, \apj, 499, 758
\reference{KenyonHartmann87}
	Kenyon,  S. J., \& Hartmann, L. 1987, \apj, 322, 293
%\reference{KenyonHartmann95}
%	Kenyon,  S. J., \& Hartmann, L. 1995, \apss, 101, 117
\reference{MCM2000}
	Marcy, G. W., Cochran, W. D., \& Mayor, M. 2000, in Protorstar and Planets~IV, V. Mannin, A. P. Boss and S. R. Russell eds., The University of Arizona Press, Tucson, p. 1285
\reference{MJMG91}
	McCaughrean, M. J., \& Gezari, D. Y. 1991, in ASP Conf. Ser. 14, Astrophysics with Infrared Arrays, Ed. R. Elston (San Francisco: ASP), 301
\reference{MJM+Stauffer94}
	McCaughrean, M. J., \& Stauffer, J. R., 1994, \aj, 108, 1382
\reference{MJM+98}
	McCaughrean, M. J., Chen, H., Bally, J., Erickson, E., Thompson, R., Rieke, M., 
Schneider, G., Stolovy, S., \& Young, E. 1998, \apj, 429, L157
\reference{MSC2000}
	McCaughrean, M. J., Stapelfeldt, K. R., \& Close, L. M. 2000, in Protorstar and Planets~IV, V. Mannin, A. P. Boss and S. R. Russell eds., The University of Arizona Press, Tucson, p. 485
\reference{Natta93}
	Natta, A. 1993, ApJ, 412, 76115.	
\reference{Natta+Panagia76}
	Natta, A., \& Panagia, N. 1976, \aap, 50, 191 
%\reference{O'Dell+Wen94}
%	O'Dell, C. R., \& Wen, Z. 1994, ApJ 436, 194
%\reference{O'Dell+93}
%	O'Dell, C. R., Wen, Z., \& Hu, XX 1993, ApJ 410, 696
\reference{Osterbrock89}
	Osterbrock, D. E. 1989, Astrophysics of Gaseous Nebulae and Active Galactic Nuclei, University Science Books, Mill Valley, California
\reference{O'Dell2001a}
	O'Dell, C. R. 2000, \araa, 39, 990 (2001a)
\reference{O'Dell2001b}
	O'Dell, C. R. 2000, \pasp, 113,209 (2001b)
\reference{O'DellYus2000}
	O'Dell, C. R., \& Yusef-Zadeh, F. 2000, \aj 120, 382
%\reference{O'Dell+93}
%	O'Dell, C. R., Wen, Z., \& Hu, XX 1993, ApJ 410, 696
\reference{PS99}
	Palla, F. \& Stahler S. W. 1999, \apj, 525, 772
\reference{Panagia73}
	Panagia, N. 1973, \aj, 78, 929
\reference{Panagia74}
	Panagia, N. 1974, \apj, 192, 221
\reference{Panagia78}
	Panagia, N. 1978, in Infrared Astronomy, 
ed. G. Setti and F. Fazio, D. Reidel Publ. Co., p. 115
\reference{Pastor+91}
	Pastor, J., Cant\'o, J., \& Rodriguez, L. F. 1991, \aap, 246, 551
\reference{Rebull2001}
	Rebull, L. M. 2001, \aj, 121, 1676
%\reference{Richling+Yorke2000}
%	Richling, S., \& Yorke, H. W. 2000, \apj 539, 258
\reference{Robberto+2002}
	Robberto, M., Beckwith, S. V. W., Herbst, T. M. 2002, in prep.
\reference{RudenPollack91}
	Ruden, S. P., \& Pollack, J. B. 1991, \apj, 375, 740
\reference{ShakuraSunyaev73}
	Shakura, N. I., \& Sunyaev, R. A. 1973, \aap, 24, 337 
\reference{SAL87}
	Shu, F., Adams, F. C., \& Lizano, S. 1987, \araa, 25,23
\reference{SMMV99}
	Stassun, K. G., Mathieu, R. D., Mazeh, T., \& Vrba, F. J. 1999, \aj, 117, 2941 
\reference{St\"orzer+Hollenbach1999}
	St\"orzer, H., \& Hollenbach, D. 1999, \apj, 515, 669
\end{references}
\end{document}